\documentclass[sigconf]{acmart}
\usepackage{multirow}
\usepackage{enumitem}
\usepackage{moreenum}
\usepackage{subfloat}
\usepackage{subfig}
\usepackage{stfloats}
\usepackage{enumitem}
\usepackage{balance}
\usepackage[title]{appendix}
\setitemize{noitemsep,topsep=0pt,parsep=0pt,partopsep=0pt}
\newtheorem{theorem}{Theorem}
\newtheorem{lemma}{Lemma}
\newtheorem{example}{Example}
\newtheorem{definition}{Definition}

\def\BibTeX{{\rm B\kern-.05em{\sc i\kern-.025em b}\kern-.08emT\kern-.1667em\lower.7ex\hbox{E}\kern-.125emX}}

\copyrightyear{2022}
\acmYear{2022}
\setcopyright{acmcopyright}
\acmConference[CCS '22]{Proceedings of the 2022 ACM SIGSAC Conference on Computer and Communications Security}{November 7--11, 2022}{Los Angeles, CA, USA}
\acmBooktitle{Proceedings of the 2022 ACM SIGSAC Conference on Computer and Communications Security (CCS '22), November 7--11, 2022, Los Angeles, CA, USA}
\acmPrice{15.00}
\acmDOI{10.1145/3548606.3560615}
\acmISBN{978-1-4503-9450-5/22/11}

\begin{document}

%
\title{Phishing URL Detection: \\A Network-based Approach Robust to Evasion}
\renewcommand{\shorttitle}{Phishing URL Detection: A Network-based Approach Robust to Evasion}

\author{Taeri Kim}
\authornote{Two first authors have contributed equally to this work.}
\affiliation{
	\institution{Hanyang University}
	\city{Seoul}
  	\country{Korea}
}
\email{taerik@hanyang.ac.kr}

\author{Noseong Park}
\authornotemark[1]
\affiliation{
	\institution{Yonsei University}
	\city{Seoul}
  	\country{Korea}
}
\email{noseong@yonsei.ac.kr}

\author{Jiwon Hong}
\affiliation{
	\institution{Hanyang University}
	\city{Seoul}
  	\country{Korea}
}
\email{nowiz@hanyang.ac.kr}

\author{Sang-Wook Kim}
\authornote{Corresponding author.}
\affiliation{
	\institution{Hanyang University}
	\city{Seoul}
  	\country{Korea}
}
\email{wook@hanyang.ac.kr}

%
\renewcommand{\shortauthors}{Kim et al.}

%
\begin{abstract}
Many cyberattacks start with disseminating phishing URLs. When clicking these phishing URLs, the victim's private information is leaked to the attacker. There have been proposed several machine learning methods to detect phishing URLs. However, it still remains under-explored to detect phishing URLs with \emph{evasion}, \textit{\textit{i.e.,}} phishing URLs that pretend to be benign by manipulating patterns. In many cases, the attacker i) reuses prepared phishing web pages because making a completely brand-new set costs non-trivial expenses, ii) prefers hosting companies that do not require private information and are cheaper than others, iii) prefers shared hosting for cost efficiency, and iv) sometimes uses benign domains, IP addresses, and URL string patterns to evade existing detection methods. Inspired by those behavioral characteristics, we present a \textit{network-based inference} method to accurately detect phishing URLs camouflaged with legitimate patterns, \textit{i.e.,} robust to evasion. In the network approach, a phishing URL will be still identified as \textit{phishy} even after evasion unless a majority of its neighbors in the network are evaded at the same time. Our method consistently shows better detection performance throughout various experimental tests than state-of-the-art methods, \textit{e.g.,} F-1 of 0.89 for our method vs. 0.84 for the best feature-based method.
\end{abstract}

%
%


%

%

%
\maketitle

\section{Introduction}
Cyberattacks cause huge damage to our society. Many cyberattacks start with phishing. Phishing is to trick people into revealing their sensitive information to the attacker. In particular, phishing URLs are camouflaged as URLs that look familiar to people. Careless people will click them, causing their private information to be leaked. Therefore, many detection methods have been developed and as a response, attackers started to consider evasion techniques that camouflage with legitimate patterns (see Section~\ref{sec:eva} for more details)~\cite{Oliveira:2017:DSP:3025453.3025831,Lin:2019:SSE:3349608.3336141,Ho:2019:DCL:3361338.3361427,adv, Ehab}. Thus, it is of utmost importance to prevent phishing attacks using evasion.

There have been proposed machine learning methods to detect phishing. They can be categorized into two types: content-based and URL string-based. \textit{Content-based methods} download and analyze web page contents~\cite{8015116,DBLP:conf/icitst/MohammadTM12,2017arXiv170107179S}. However, they require non-trivial computations to process many web pages and are weak against web browser-based exploits (because we need to access their web pages). Most importantly, it is not easy to collect such training data. For all those reasons, content-based methods are not always preferred. \textit{String-based methods} mainly rely on URL string pattern analyses because it is well known that phishing URLs have very distinguishable string patterns~\cite{Ma09beyondblacklists,Blum:2010:LFB:1866423.1866434,6061361,DBLP:conf/icitst/MohammadTM12,Mohammad2014,7207281,Verma:2015:CPU:2699026.2699115,7945048,2017arXiv170107179S,hong2020phishing,anand2018phishing}. Thus, many lexical features to detect phishing URLs have been proposed (see Section~\ref{sec:rel}). These features are known to be effective in detecting phishing URLs. Because string-based methods are computationally lightweight and provide high accuracy, many researchers prefer them for the high efficiency~\cite{2017arXiv170107179S}. Some researchers rely on a blacklist of IP addresses and domains. However, its accuracy is known to be mediocre.

Almost all existing string-based methods hardly consider evasion~\cite{adv}. Evasion means the technique that the attacker creates phishing URLs seemingly legitimate by manipulating their patterns to deceive defenders' detection methods. In this work, we consider a couple of more key patterns of phishing attacks to design an advanced string-based detection method that outperforms existing methods and is strong against evasion. First, the attacker is sensitive to cost efficiency~\cite{apwg}. In many cases, they (partially) reuse phishing attack materials and prefer specific hosting companies for their looser policies (\textit{e.g.,} not to require identification information) and relatively cheaper prices than other agencies. When a private server is used instead of hosting companies, the attacker prefers shared hosting, \textit{i.e.,} one server is used for multiple phishing attack campaigns and also for multiple domains --- in our data, 15.8\% of IP addresses are connected to multiple domains. Second, the attacker creates phishing URLs on top of benign servers, domains, IP addresses, and/or substrings to evade existing detection methods~\cite{apwg}.

\begin{figure*}
\centering
\includegraphics[width=0.99\textwidth]{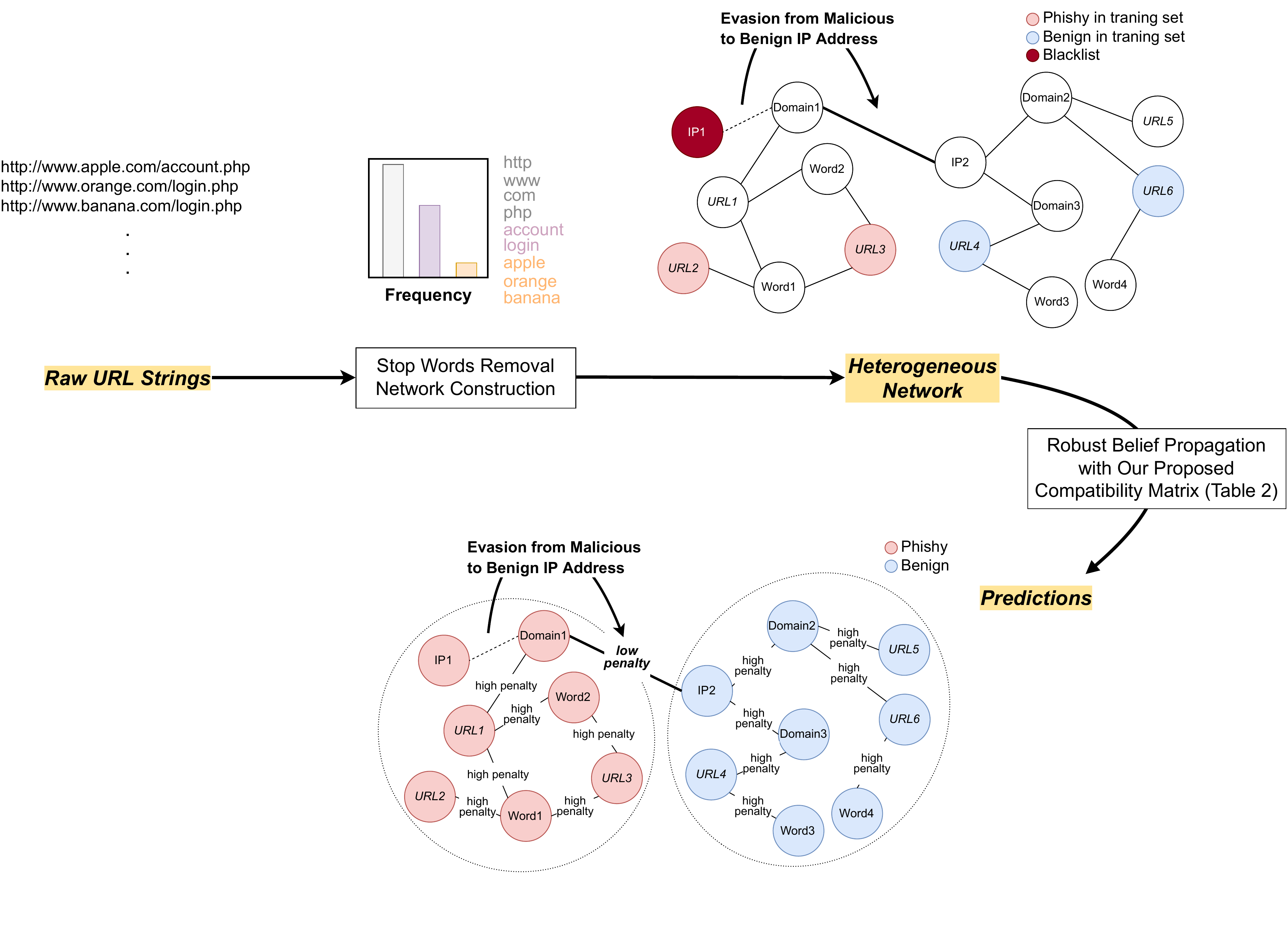}
\vspace{-3em}
\caption{The overall workflow of the proposed method. In the first step, we segment collected URLs into words and remove meaningless ones that correspond to stop words that have high frequency but do not carry useful information. In the second step, we construct a heterogeneous network of URLs, Domains, IP addresses, etc. In the last step, we run the customized belief propagation method to make it robust.}
\label{fig:approach}
\vspace{-1em}
\end{figure*}

Considering all these facts, we design a novel unified framework of natural language processing and a network-based approach to detect phishing URLs --- its overall workflow is shown in Fig.~\ref{fig:approach}. We regard each URL as a sentence and segment it into substrings (words) considering the syntax and punctuation symbols of URLs --- URLs have well defined syntax as in English. After that, we build one large network that consists of heterogeneous entities, such as URLs, domains, IP addresses, authoritative name servers, and substrings, and perform our \emph{customized belief propagation} to detect phishing URLs (see Section~\ref{edge_potential}). We note that the above listed related works do no include any network-based inference schemes. On the contrary, similar network-based inference methods had been used in various other domains~\cite{Manadhata2014,chau2011polonium}. However, our method differs from them in defining \emph{edge potentials} which decide a penalty when two neighboring entities have different predicted labels.

Our approach is effective to infer that seemingly unrelated phishing URLs are actually related and is robust to evasion. Because we infer on such a network of heterogeneous entities, \emph{an evasion for a phishing URL is not likely to be successful unless a majority of its neighbors in the network are evaded at the same time} (see Section~\ref{sec:rob} for more detailed discussions with theorems and proofs), which is our main contribution in comparison with existing works.

We crawled many suspicious URLs and also downloaded a couple of datasets released by other researchers~\cite{Sorio2013DetectionOH,ahmad}. In total, we have about 120K phishy and 380K benign URLs. We compare our approach with state-of-the-art baseline methods including graph convolutional networks (GCNs) and feature engineering-based methods. Our method shows the best detection performance among them. Furthermore, in additional evasion tests, our method shows better F-1 scores than other baseline methods. Because the evasion incurs non-trivial expenses for the attacker to access to benign domains, IP addresses, and so forth, our robust detection method greatly increases the attacker's financial burden to perform evasion.

Our contributions can be summarized as follows:
\begin{itemize}
    \item We design a novel network-based inference method equipped with our proposed robust edge potential assignment mechanism. Our network inference on top of the edge potential assignment outperforms many baseline methods including feature engineering-based and network-based classifiers.
    \item Our proposed network-based method has a theoretical ground on why it is robust to evasion (see Section~\ref{sec:rob}).
    \item We conduct experiments with a large set of URLs collected by us and downloaded from other work. Our data covers a wide variety of phishy/benign URL patterns.
\end{itemize}

In the following, we first review the literature in Section~\ref{sec:rel} and describe the motivation of this work in Section~\ref{sec:eva}. Then, in Sections~\ref{sec:proposedMethod} and ~\ref{sec:rob}, we design a novel network-based detection method robust to evasion and analyze the theoretical robustness of the proposed method. After that, we conduct extensive experiments on phishing URL detection with and without evasion in Section~\ref{sec:Experiments}. Lastly, in Sections~\ref{sec:crawl} and~\ref{sec:conclusions}, we describe crawled our data and conclude our paper.
For reference, in Appendix~\ref{sec:baseline}, we introduce a set of lexical features widely used to detect phishing URLs and sorted in descending order of the feature importance extracted from the best performing baseline method.

\section{Related Work}\label{sec:rel}
In this section, we review phishing URL detection models and attackers' behavioral pattern analyses.

\subsection{Methods to Detect Phishing URLs}Extensive work has been done to counter phishing attacks~\cite{Ma09beyondblacklists,Blum:2010:LFB:1866423.1866434,6061361,DBLP:conf/icitst/MohammadTM12,Mohammad2014,7207281,Verma:2015:CPU:2699026.2699115,7945048,8015116,2017arXiv170107179S}. Typically, researchers have explored machine learning techniques to automatically detect phishing URLs. It is vital to have a well-defined set of features for the effectiveness of classification algorithms. So, we introduce a widely used set of 19 URL features that we collected from related papers in Appendix~\ref{sec:baseline}. All these features are used by some baseline methods in our experiments. All the mentioned works are not based on network-based inference but on feature engineering.

Mao et al. designed a phishing URL detection method robust to evasion based on web page content features~\cite{8015116}. However, it is not easy to collect such training data in many cases because phishing attacks do not last long and web pages are quickly removed, which is one common drawback of all content-based detection methods~\cite{6061361}.

In~\cite{7945048,2018arXiv180203162L,melissa-dl}, several sequence (\textit{e.g.,} URL in our context) classification models have been proposed. Some of them have an advanced architecture to combine various components such as recurrent neural networks, convolutional neural networks, word embeddings, and their multiple hierarchical layers. We use their ideas as additional baselines. The first one uses long short-term memory (LSTM) cells and the second model uses one-dimensional convolution (1DConv), and the third baseline uses both (1DConv+LSTM).

For a couple of related problems~\cite{Manadhata2014,chau2011polonium}, network-based methods have been used. In~\cite{Manadhata2014}, the authors tried to detect malicious domains (rather than URLs) and the authors in~\cite{chau2011polonium} proposed one heuristic-based belief propagation method to detect malicious codes. Those two works differ in how to create networks but use the same belief propagation method. Both methods correspond to the baseline method marked as `POL' in our experiments. Peng et al. and Khalil et al. also tried a network approach for malicious domain detection~\cite{10.1007/978-3-030-12981-1_34,Khalil:2018:DOG:3176258.3176329}. However, their methods are not directly applicable to our phishing URL data.

\subsection{Attackers' Behavioral Patterns}\label{sec:attacker}
Phishing Activity Trends Report~\cite{apwg} by Anti-Phishing Working Group is one of the most reputable reports. We analyzed their quarterly reports. The two most important observations from the reports are i) there are some web hosting companies preferred by the attacker due to their low prices and anonymity, and ii) many phishing URLs have similar string patterns because they are created by similar tools or reused from old phishing campaigns. There exist many other interesting observations as follows:
\begin{itemize}
    \item There has been an increase in the number of phishings using free hosting providers or website builders. It has been reported that 81.7\% of malicious websites are hosted on free hosting providers~\cite{de2021compromised}. These free hosts are easy to use but also allow threat actors to create subdomains spoofing a targeted brand, resulting in a more legitimate-looking phishing site. Free hosts also afford phishers additional anonymity, because these services hide registrant information.
    \item The attacker prefers shared hosting which means multiple domains share the same hosting server. Therefore, seemingly unrelated domains may belong to the same host or IP address.
    \item Hundreds of vendors are mostly targeted. This continues a years-long trend in which a few hundred companies are attacked regularly. Considering this fact, we crawled URLs from \url{phishtank.com} for the three most frequently attacked vendors: Bank of America, eBay, and PayPal.
    \item 53\% of phishing attacks use `com' domains and `net', `org', and `br' domains are next equally preferred.
\end{itemize}

\section{Motivation}\label{sec:eva}
\begin{definition}
Evasion is an effective technique that one can adopt to disturb a machine learning task by creating a `counter-evident' sample, \textit{e.g.,} a phishing URL hosted by a benign domain or IP address. This evasion can be done in various ways. For detailed evasion techniques that we consider, refer to Section~\ref{sec:evasion}.
\end{definition}
\vspace{-6px}
Shirazi et al. showed that existing phishing URL detection methods are adversely impacted by evasion without suggesting a countermeasure~\cite{adv}.
Specifically, they conducted evasion tests that randomly select up to four features of phishing URLs and change the selected features to other benign values. In their non-evasion tests, most classifiers showed high accuracy. In their evasion tests, however, the best performing classifier's accuracy (recall) decreased from 82-97\% to 79-45\% with one feature change, and to 0\% with four feature changes.

To our knowledge, it has not been actively studied to design a non-content-based phishing URL detection method robust to evasion. We consider many aspects of URLs, including domains, IP addresses, name servers, and string patterns \emph{except contents} --- because collecting phishing web page contents require non-trivial efforts. Most importantly, our method is based on a network of them. Intuitively speaking, attackers cannot disturb our network-based inference task even after evasion if many neighbors of a phishing URL in the network remain the same as before (see Section~\ref{sec:rob}). Some large-scale evasion can still neutralize our method. However, it requires non-trivial expenses, thus decreasing the attackers' motivation on such evasion.

While it is hard to measure the evasion cost for money, it includes various intangible efforts, such as exploiting benign web servers to implant their phishing pages, maintaining a custom domain without any phishing campaigns until D-Day to prevent it from being blacklisted, and so forth. In particular, it depends on security environments and skills how long it will take until an attacker successfully exploits an administrator's account of a benign server.

\section{Proposed Method}\label{sec:proposedMethod}
After introducing the overall workflow in our method, we describe its detailed steps with some key visualization results.

\subsection{Overall Method}Fig.~\ref{fig:approach} shows our overall workflow. The entire process can be divided into the following steps:
\begin{enumerate}
    \item We crawl many URLs from \url{phishtank.com} and download other works' open datasets.
    \item As mentioned earlier, we create a heterogeneous network of URLs, domains, IP addresses, name servers, and substrings (words). We use a standard natural language processing technique to segment URLs into substrings (words) and draw edges between a URL and substrings.
    \item We run our customized belief propagation algorithm to infer unknown URLs' phishy/benign labels, which is our main contribution. In particular, this type of inference is called \emph{transductive}. In our case, both training and testing samples co-exist in a network and testing samples' labels are inferred from other known training samples' labels following the network architecture. 
\end{enumerate}

\begin{figure}
\centering
\includegraphics[width=0.7\columnwidth,trim=1.8 0 0 0.71cm, clip]{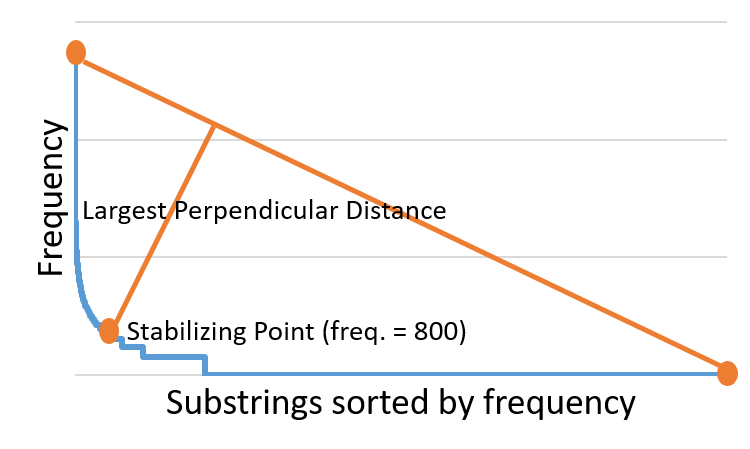}
\vspace{-1.5em}
\caption{The elbow method to find the stabilizing point of frequency. All substrings (words) before the found stabilizing point are considered as stop words.}
\label{fig:elbow}
\end{figure}

\subsection{Network Construction}\label{sec:ne}
We do a network-based classification rather than feature engineering-based classification. As mentioned earlier, phishing URLs share many common string patterns and various entities are cross-related to each other, so we create a network to represent complicated relationships among multiple entities (vertices) such as URLs, their domains, IP addresses, authoritative name servers, and substrings.
\begin{itemize}
    \item We draw an edge between a URL and its domain.
    \item We draw an edge between a domain and its resolved IP address. We use \url{domains.google} and \url{virustotal.com} to retrieve domain-IP address resolution history. They return not only current but also all past resolution results with timestamps which enable correct connections. Sometimes, one domain can be connected to multiple IP addresses.
    \item We draw an edge between a domain and its authoritative name servers. In general, there exist multiple authoritative name servers for a domain, and one authoritative name server provides resolution services for multiple domains.
    \item We draw an edge between a URL (\textit{i.e.,} sentence) and a substring (\textit{i.e.,} word) if the URL contains the substring. For these edges, it is very crucial how to segment a URL into substrings. We will shortly describe this in the following section.
\end{itemize}

\begin{figure}
\centering
\includegraphics[width=0.99\columnwidth,trim=0 0.95 0 0.1cm, clip]{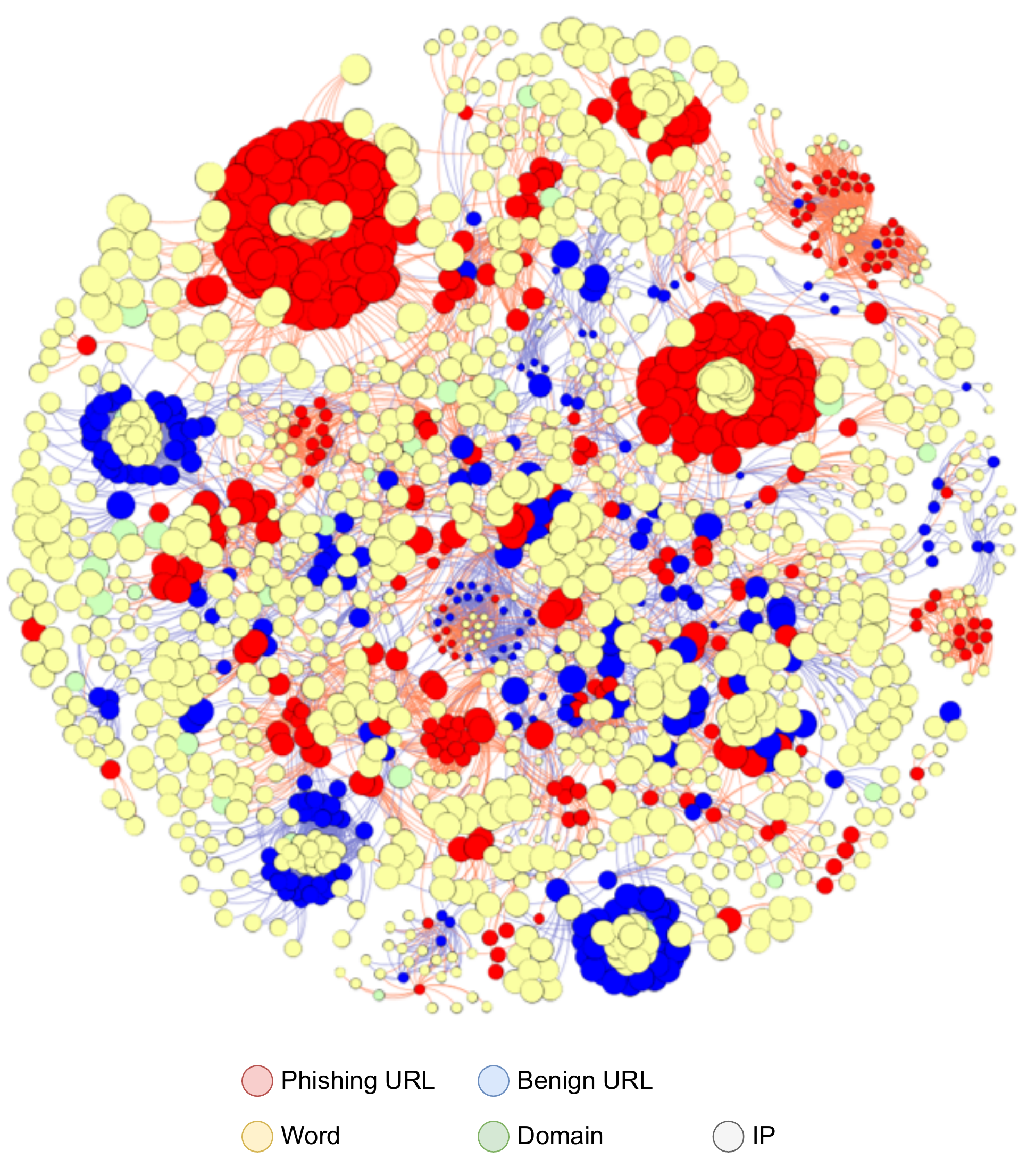}
\vspace{-0.8em}
\caption{The network constructed from our data. Red means phishing URLs and blue means benign URLs. Other colors mean non-URL entities --- name servers are not displayed due to their lesser significance than that of other entities. Note that there exist many clusters. The vertex size represents the strength (more specifically, modularity~\cite{Blondel_2008}) of the cluster that a vertex belongs to.}
\label{fig:network}
\end{figure}

\subsubsection{How to segment a URL into words}\label{sec:seg}
A URL is used to locate resources in the Internet. It consists of several parts: scheme, username, password, host, port number, path, and query string --- some of them can be missing. We use our customized word segmentation policies in each part as follows:
\begin{itemize}
\vspace{0.5em}
    \item \textit{Scheme} means the protocol, \textit{e.g.,} http and https. Only two words can be possible. However, since these words have very high frequencies, we do not use these two words in our network. We will describe how to remove those \emph{stop words}\footnote{Stop words do not carry meaning but have high frequency values in English, such as `a', `the', `is', and so forth. It is a standard process to remove such stop words in natural language processing algorithms. We use the elbow method based on frequency to detect the stop words of URLs.} of URLs shortly.
    \item \textit{Username} and \textit{password} can be specified before host. We segment them using the punctuation symbols, \textit{i.e.,} `//', `:', and `@'. An example is `http://username:password@example.com'.
    \item \textit{Hostname} can be simply segmented into words by `.'.
    \item Sometimes \textit{path} can be very long, separated by `/'. We use all possible punctuation symbols, such as `/', `.', `!', `\&', `,', `\#', `\$', `\%', and `;', to segment the path part into words.
    \item \textit{Query string} is able to contain multiple queries separated by `\&', and each query consists of a query name and a value, \textit{e.g.,} `term=bluebird\&source=browser-search'. We extract words using the two punctuation symbols, `=' and `\&'.
\end{itemize}
\vspace{0.5em}

Because the syntax of URLs is well defined, extracting words can be done very efficiently. However, many meaningless words can be also extracted. Therefore, before drawing edges between URLs and words, those words should be removed. In the field of natural language processing, it is well known that the frequency of words follows Zipf's law --- more precisely, word frequency exponentially decays~\cite{33858}. In particular, this pattern describes stop words in English very well. For instance, the frequency of the most popular stop word `the' occupies 7\% of all word occurrences in the Brown Corpus of American English~\cite{francis79browncorpus} and the second most popular stop word `of' has 3.5\%. We found that the extracted words from URLs show similar statistics (cf. Fig.~\ref{fig:elbow}). After that, we remove some high-frequency words using the \textit{elbow method}~\cite{ketchen1996application}. It decides the point whose perpendicular distance to the line segment connecting the two ends is the biggest as the saturation point, which is 800 in our data. We remove all the words whose frequency values are larger than the point.

Fig.~\ref{fig:network} shows the network created by the proposed method. Note that there exists strong correlation between the cluster constructions and the ground-truth phishy/benign labels, which justifies our network-based inference method that will be described shortly. \emph{In this regard, the main intuition in our work is that it is hard to evade our natural language processing and network-based approach unless a majority of entities in a cluster are evaded simultaneously.}

\subsection{Network-based Inference} We employ \textit{loopy belief propagation} (LBP)~\cite{Bishop:2006:PRM:1162264} for our network-based inference. Our key contribution in this step is to define a more advanced edge potential assignment mechanism than that of the state-of-the-art methods~\cite{chau2011polonium,Manadhata2014}.
Because these methods typically not only follow a majority voting of neighbors but also give a fixed edge potential regardless of the similarity of the two connected vertices, a vertex is mainly classified as benign if it has many benign neighbors.
However, we want to correctly classify a phishing vertex even if it has many benign neighbors. Therefore, we define a more advanced edge potential assignment mechanism for enabling more sophisticated classification and achieving evasion-robustness.
We will describe our edge potential definition in Section~\ref{edge_potential}.

LBP is a message passing algorithm to solve network-based inference problems. Let $x \in X$ be a hidden variable and $N_x$ be a set of its neighboring variables, and let $o \in O$ be an observed variable. In our contexts, an observed variable means a training sample and a hidden variable means a testing sample. We use $X$ and $O$ to denote a set of all hidden and observed variables, respectively. Each variable represents the phishy/benign label of an entity in our case. $x$ sends a message to other hidden variable $y \in N_x$ after collecting all messages from $N_x \setminus \{y\}$. Note that observed variables never receive any messages; they only broadcast the messages to their neighboring hidden variables. In our case, phishy and benign URLs in the training set are observed variables.

As mentioned, we need to calculate a message $msg_{x \rightarrow y}(\ell)$ from a variable $x$ to other variable $y$ regarding a phishy/benign label $\ell \in L$, where $L=\{phishy, benign\}$ is a set of all possible label options. There exist several message passing strategies: \textit{sum-product}, \textit{max-product}, and \textit{min-sum}. We use the min-sum algorithm having better computational stability than the other two algorithms. For some high degree vertices, message values tend to quickly decay to zeros (\textit{i.e.,} floating point underflow) in the sum-product and max-product. Their product operation is reduced to the sum in the min-sum algorithm.
The message in the min-sum algorithm is calculated as:
\vspace{0.4em}
\begin{align}\label{eq:msg}
\begin{split}
msg_{x \rightarrow y}(\ell) = &\min_{\ell'} \Big[ \log{\left(1 - \phi_y(\ell')\right)} + \psi_{xy}(\ell, \ell') + \\ 
&\sum_{k \in N_x \setminus \{y\}} msg_{k \rightarrow x}(\ell') \Big],
\end{split}
\end{align}
\vspace{0.5em}
where $\phi_y(\ell')$ is a \textit{prior} that the variable $y$ has the label $\ell'$ and $\psi_{xy}(\ell, \ell')$ is an \textit{edge potential}, indicating a joint-probability that $x$'s label is $\ell$ and $y$'s label is $\ell'$. Note that there is a log function in the message definition so the min-sum is equivalent to performing the max-product in the log space for better computational stability.

After exchanging messages many times, we first calculate a \textit{cost} of each variable and label pair and then choose the label that yields the \textit{lowest cost}\footnote{The min-sum tries to minimize `cost' as the name `min' suggests whereas both the sum-product and max-product maximize `belief'.} for each variable. The cost, when $x$ has the label $\ell$, is computed as:
\vspace{0.3em}
\begin{align}\label{eq:cost}
Cost(x,\ell) = \log{\left(1 - \phi_x(\ell)\right)} + \sum_{k \in N_x} msg_{k \rightarrow x}(\ell).
\end{align}

Then, the formal definition of the problem that the min-sum algorithm solves can be defined as follows:
\vspace{0.4em}
\begin{align}\label{eq:minsum}
\operatorname{argmin}_{g} \sum_{x}Cost(x, g(x)),
\end{align}
\vspace{0.4em}
where $g: X \rightarrow L$, where $X$ is a set of hidden variables and $L=\{phishy, benign\}$, is a label assignment function. It is worth mentioning in our setting, $x$ can be a hidden variable representing a URL, domain, IP, name server, or word. Our final target is to infer the labels of testing URLs. To this end, we need to infer the labels of other non-URL entities as well because they connect URLs. Therefore, the min-sum algorithm can be described as a process of finding such label assignments to hidden variables that the sum of the costs is minimized.
 
\subsubsection{Edge Potential Assignment} \label{edge_potential}
The definition of edge potential $\psi_{xy}(\ell, \ell')$ is the key factor in the LBP method.~\cite{chau2011polonium} used the heuristics of \textit{homophily} and \textit{heterophily}. They, for example, assign an edge potential of $0.5 - \epsilon$ (resp. $0.5 + \epsilon$) if two neighboring variables $x$ and $y$ have different (resp. same) labels as shown in the compatibility matrix in Table~\ref{table:polonium}. $\epsilon$ is usually set as very small, \textit{e.g.,} 0.001. We use two labels, phishy and benign. At the end of the network-based inference process, for each entity, one label will be assigned as a prediction result. The final label assignments are greatly influenced by the edge potential definition.

In contrast to~\cite{chau2011polonium}, we incorporate more factors, such as similarity among entities and an improved compatibility matrix, to derive reliable edge potentials --- we prove shortly in Section~\ref{sec:rob} that reliable similarity definitions can lead to the evasion-robustness in our method. The similarity can be measured via various embedding approaches, such as Doc2Vec \cite{le2014distributed} and Node2Vec \cite{grover2016node2vec}. We discuss how to calculate vector representations of URLs, their domains, IP addresses, authoritative name servers, and words in Section~\ref{embedding}.

\begin{table}[t!]
\centering
\caption{The compatibility matrix proposed in Polonium~\cite{chau2011polonium} based on the homophily heuristic}\label{table:polonium}
\vspace{-0.8em}
{\renewcommand{\arraystretch}{1.3}
  \begin{tabular}{c | c  c}
  $\psi_{xy}({\ell, \ell'})$ & Phishy & Benign \\ [0.5ex]
  \hline
  Phishy & $0.5 + \epsilon$ & $0.5 - \epsilon$ \\
 Benign & $0.5 - \epsilon$ & $0.5 + \epsilon$ \\
  \end{tabular}

\vspace{1.0em}

\caption{Our compatibility matrix $M$ for the min-sum algorithm. $\mathbf{x}$ and $\mathbf{y}$ mean vector representations. $sim(\mathbf{x},\mathbf{y})$ is a similarity between two vectors.}
\label{table:2}
\vspace{-0.8em}
  \resizebox{\columnwidth}{!}{\begin{tabular}{c | c  c}
  $\psi_{xy}({\ell, \ell'})$ & Phishy & Benign \\ [0.5ex]
  \hline
  Phishy & $\min(ths_+, 1-sim(\mathbf{x}, \mathbf{y}))$ & $\max(ths_-, sim(\mathbf{x}, \mathbf{y}))$ \\
  Benign & $\max(ths_-, sim(\mathbf{x}, \mathbf{y}))$ & $\min(ths_+, 1-sim(\mathbf{x}, \mathbf{y}))$ \\
  \end{tabular}}
}
\end{table}

To calculate the similarity based on those vector representations, we adopt several different similarity measures, including the cosine similarity and various kernels. Our proposed definition of edge potential is shown in Table~\ref{table:2}. In the table, we denote vector representations of entities in boldface and $sim(\mathbf{x},\mathbf{y})$ indicates a similarity between two vectors that can be defined in various ways. Two such examples are as follows:
\vspace{0.3em}
\begin{align*}
 sim(\mathbf{x},\mathbf{y}) = 
 \vspace{0.3em}
 \begin{cases}
 \vspace{0.3em}
 cos(\mathbf{x},\mathbf{y})\textrm{ based on the cosine similarity}, \\
 \exp(\frac{\|\mathbf{x}-\mathbf{y}\|^2}{2\sigma^2})\textrm{ based on the RBF kernel}.
 \end{cases}
 \vspace{10em}
\end{align*}
After that, we use a concept inspired by the \emph{hinge-loss}~\cite{Rosasco:2004:LFS:996933.996940} to assign edge potential values. For instance, $\min(ths_+, 1-sim(\mathbf{x}, \mathbf{y}))$ in the table is to limit the minimum edge potential to $ths_+$\footnote{This means a lower-bound of edge potential and is set by a user.} when two entities have the same label. When $sim(\mathbf{x}, \mathbf{y})$ is low (resp. high), the proposed definition imposes a large (resp. small) penalty closed to 1 (resp. $ths_+$). Therefore, the proposed mechanism is able to assign much more sophisticated edge potentials in comparison with existing methods.

One should be very careful when applying our compatibility matrix to other applications. Recall that we use the min-sum algorithm so that in our compatibility matrix $M$, we assign 0 (which corresponds to 1 in the sum-product and max-product algorithms) when $\ell$ and $\ell'$ are the same. For the sum-product and max-product algorithms, $1 - M$ should be used.

\begin{figure}[t]
\vspace{-1.5em}
\centering
\subfloat[Cosine Similarity]{\includegraphics[width=0.495\linewidth,trim=0 0 0 0cm, clip]{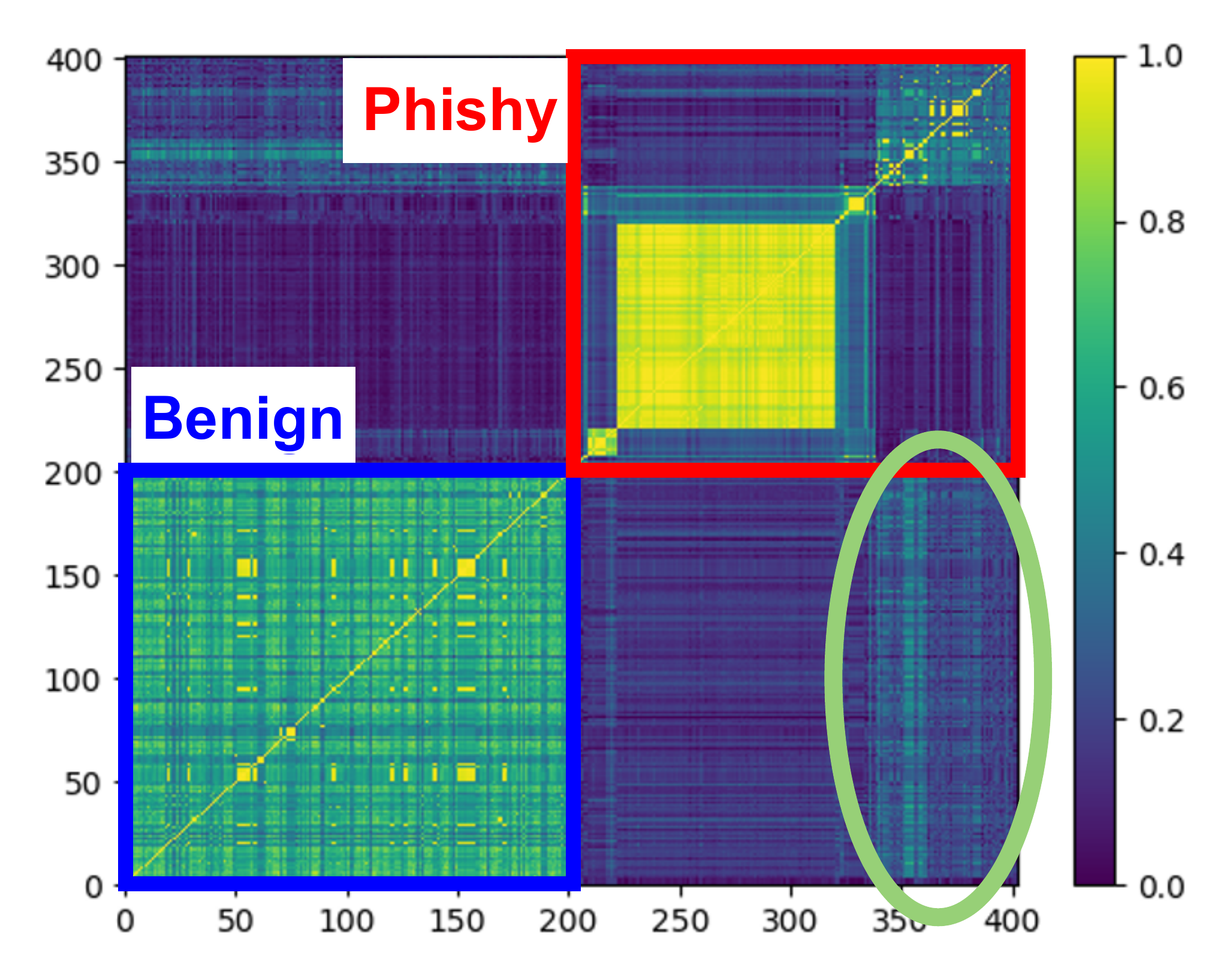}} \hfill
\subfloat[Inference Result]{\includegraphics[width=0.495\linewidth,trim=0 0 0 0cm, clip]{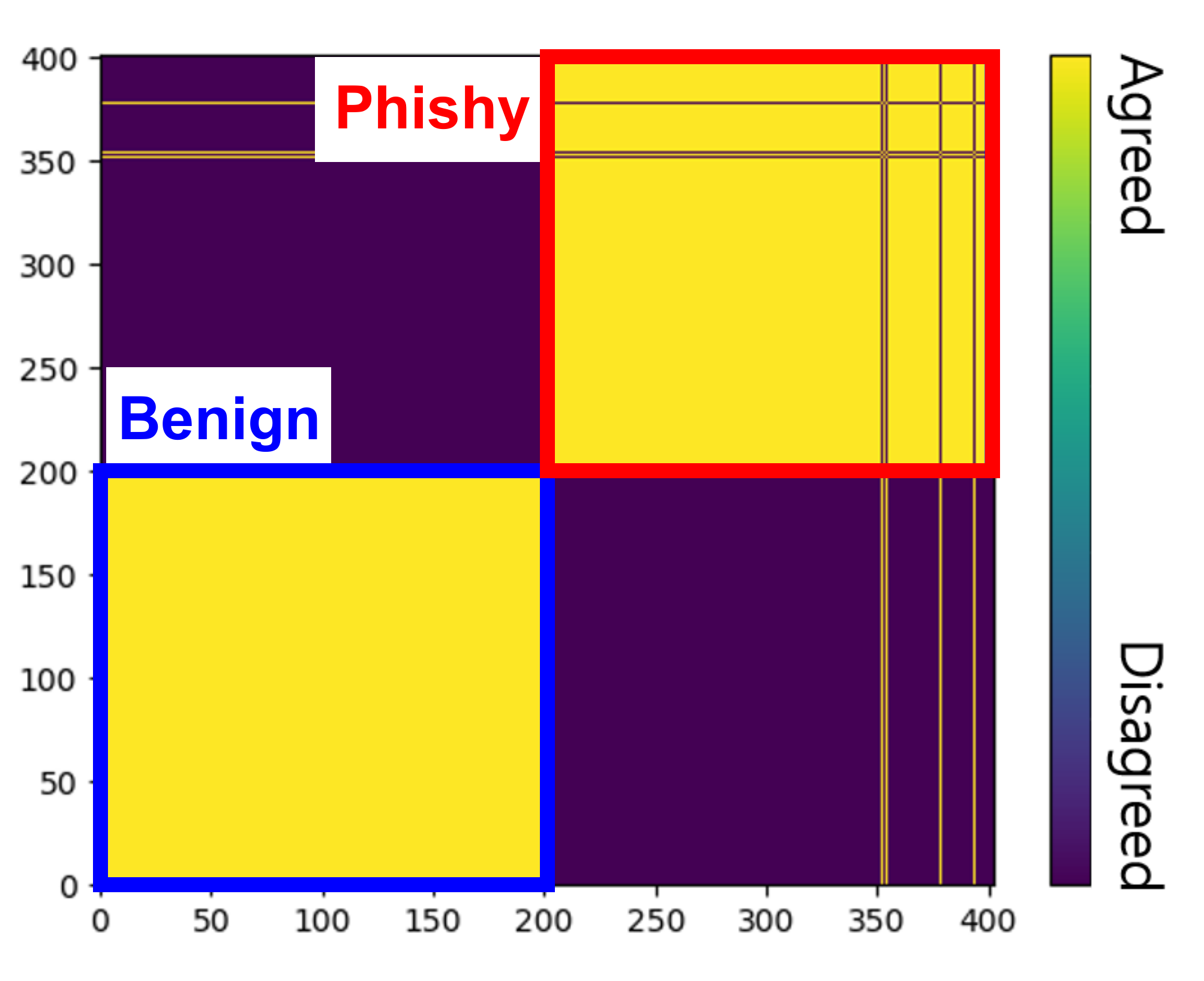}} 
\vspace{-0.8em}
\caption{Examples of the pairwise vertex similarity with DeepWalk and our network inference with $ths_+ = ths_- = 0.7$. (a) We choose the highest PageRank URL and other 199 URLs in its neighborhood with the breadth-first search for each of the phishy and benign classes. In total, there are 400 URLs in the similarity plot. (b) From the similarity, our network-based inference is able to infer almost correctly.}
\label{fig:emb}
\end{figure}

\subsubsection{Vector Representations of Entities}\label{embedding}
We describe how we can calculate reliable vector representations of various entities. These embedding methods are known to be effective in discovering latent relationships among entities~\cite{mikolov2013efficient,le2014distributed,2014arXiv1403.6652P,grover2016node2vec,yoo2022directed,lee2020asine,lee2020negative,DBLP:conf/iclr/0002JJH0JSP22}, which is a good fit to our network-based detection under the presence of evasions.

\paragraph{Word Embedding-based Methods}
In the area of natural language processing, there have been proposed various semantic embedding methods such as Word2Vec~\cite{mikolov2013efficient} and Doc2Vec~\cite{le2014distributed}. As we mentioned earlier, we segment URLs into words so we can directly apply the methods to calculate the vector representations of URLs and words. However, we cannot directly calculate vector representations of domains, IP addresses, and name servers in this approach because it considers only strings. Inspired by \textit{locally linear embedding} (LLE)~\cite{Roweis2000}, however, we propose a heuristic to represent a domain, IP address, or name server as a mean vector of its neighbors' vectors. LLE says that a vector representation of an entity is a weighted combination of its neighbors' vectors, \textit{e.g.,} equally weighted in our case. For this, we first calculate mean vector representations of domains and then IP addresses and so forth, given URLs' vector representations calculated by Word2Vec or Doc2Vec.

\paragraph{Network Embedding-based Methods}
Another reliable approach to find vector representations is to use network embedding methods. Many of these methods have been proposed by social network researchers. One advantage of the approach is that we can find vector representations of all entities simultaneously because they can run on our network directly. We use Node2Vec~\cite{grover2016node2vec} and DeepWalk~\cite{2014arXiv1403.6652P}. In Fig.~\ref{fig:emb} (a), we show a pairwise similarity plot that intuitively justifies our embedding and similarity-based edge potential assignments. However, we see a small portion of phishy and benign pairs in the green circle have high similarities. This can be corrected by our proposed edge potential assignment mechanism, which is shown in Fig.~\ref{fig:emb} (b).

\section{Evasion-Robustness of Our Network-based Approach}\label{sec:rob}
In this section, we formally prove that a hidden variable's phishy/benign label follows its \emph{similar} neighbors' majority label, which improves the robustness to evasion.

\begin{lemma}\label{lemma}
Suppose $ths_+ = ths_- = 0$ and a small network that consists of a hidden variable $u$ and its $m$ neighbors $N_u = \{v_1, \cdots, v_m\}$. Let $\ell_u$ be the phishy/benign label of $u$. When $\ell_u = \operatorname{argmin}_{\ell} \sum_j sim(\mathbf{u},\mathbf{v}_j) \cdot I(v_j, \ell)$, where $I(v_j, \ell) \in \{0,1\}$ is an indicator function saying if $v_i$ has a label $\ell$, the min-sum algorithm in Eq.~\eqref{eq:minsum} is optimized.
\end{lemma}

\begin{proof}
$\ell_u$ is inferred by Eq.~\eqref{eq:cost}. In particular, the second term in the equation, $\sum_{v \in N_u} msg_{v \rightarrow u}(\ell)$, is significant to decide its label, and $msg_{v \rightarrow u}(\ell)$ is dominated only by $\psi_{vu}(\ell_v, \ell_u)$ in the assumed network (cf. Eq.~\eqref{eq:msg}). $\sum_{v \in N_u} \psi_{vu}(\ell_v, \ell_u)$ is minimized when $\ell_u$ follows the majority label considering the vector similarities because $\psi_{vu}(\ell_v, \ell_u)$ is determined by $sim(\mathbf{v}, \mathbf{u})$ in Table~\ref{table:2}.
\end{proof}

\begin{figure}[t]
\centering
\includegraphics[width=0.7\columnwidth]{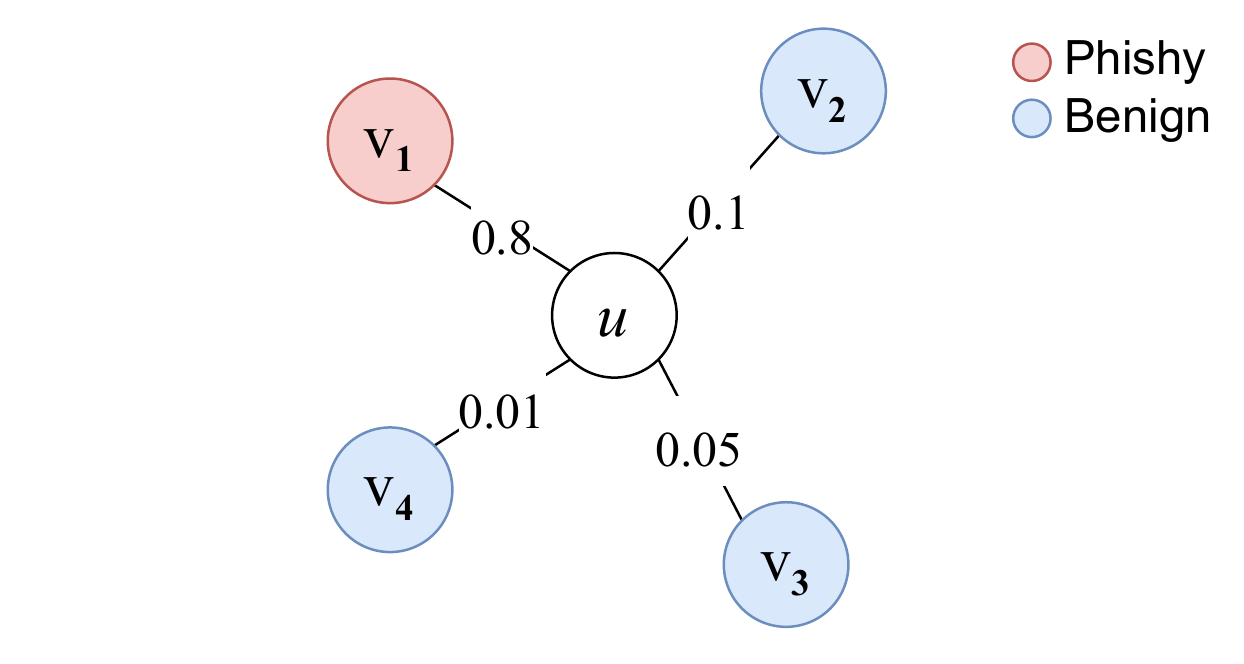}
\vspace{-0.8em}
\caption{For ease of discussion, suppose $u$ is a hidden variable and other variables' labels are fixed. Each edge is annotated with $sim(\mathbf{u}, \mathbf{v}_i)$. Our method concludes that $u$ is phishy although $u$ has more benign neighbors.}

\vspace{2em}

\label{fig:lemma}
\centering
\includegraphics[trim={1.3cm 1.8cm 1.1cm 1.2cm},clip,width=1\columnwidth]{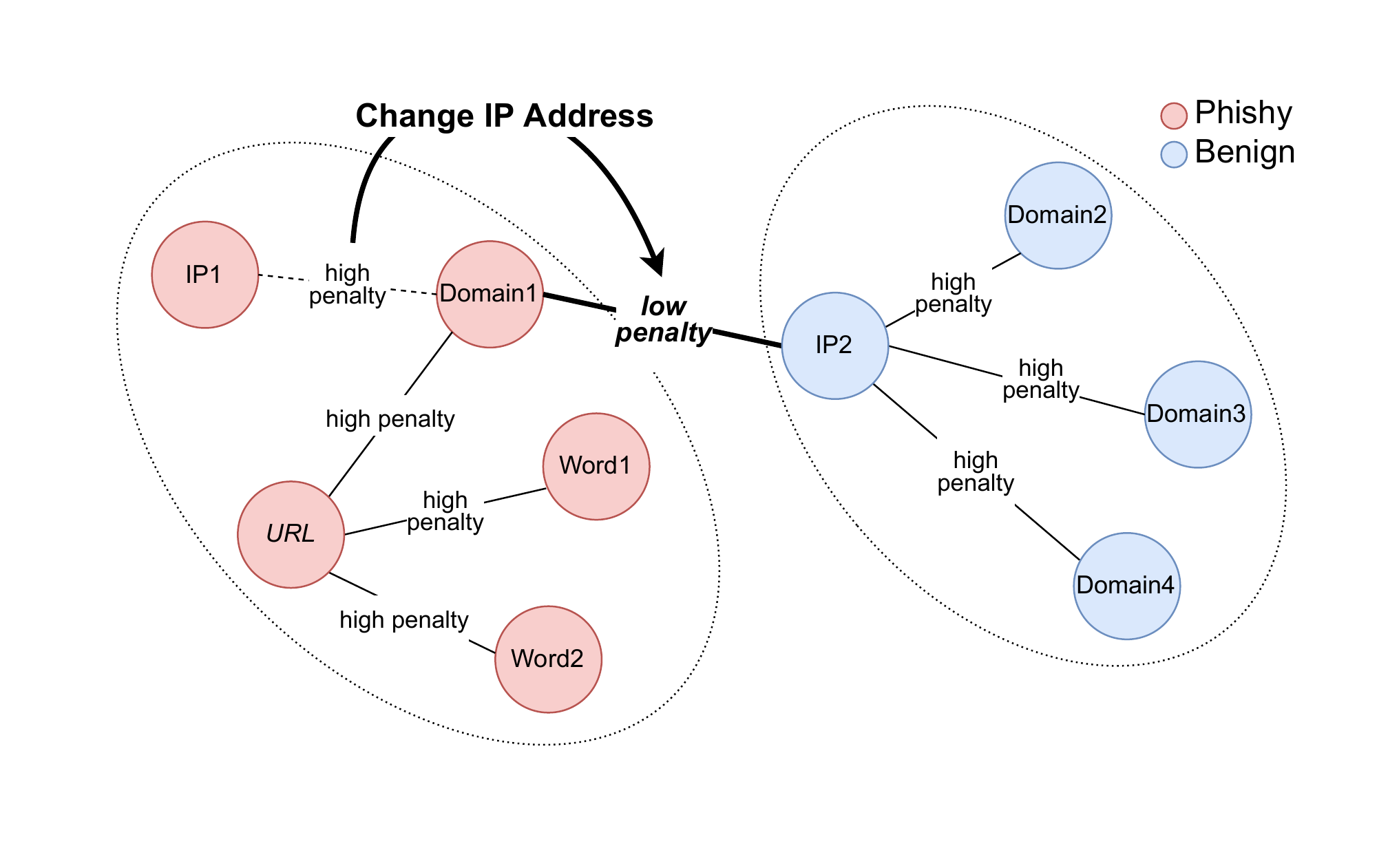}
\vspace{-1.5em}
\caption{There are two clusters. In general, connections between phishy and benign clusters are not strong (cf. Fig.~\ref{fig:network}). `Domain1' is connected to `IP2' after evasion. However, the connection between them is weak and after embedding, $sim(\mathbf{x}, \mathbf{y})$ is low, where $x=Domain1$ and $y=IP2$. Thus, a low penalty is given to their dissimilar labels by our compatibility matrix and the belief propagation can still identify `Domain1' as phishy.}
\label{fig:evasion}
\end{figure}

\begin{example}[Example of Lemma~\ref{lemma}]
In Fig.~\ref{fig:lemma}, there is the small network we used in Lemma~\ref{lemma}. For ease of discussion, suppose that only $u$ is a hidden variable and others are observed variables. The optimal min-sum solution is $g(u)=Phishy$ because $sim(\mathbf{u}, \mathbf{v}_1) > \sum_{j > 1} sim(\mathbf{u}, \mathbf{v}_j)$ and $Cost(u, Phishy) = \sum_{j > 1} sim(\mathbf{u}, \mathbf{v}_j)$ is smaller than $Cost(u, Benign) = sim(\mathbf{u}, \mathbf{v}_1)$.
\end{example}

This lemma can be generalized to the following theorem for larger general networks:

\begin{theorem}
Given a large network $G=(V,E)$, the min-sum algorithm is optimized if for each hidden variable $u \in V$ and its neighbors $N_u$, $\ell_u = \operatorname{argmin}_{\ell} \sum_j sim(\mathbf{u},\mathbf{v}_j) \cdot I(v_j, \ell)$.
\end{theorem}

\begin{proof}
 If we can achieve $\ell_u = \operatorname{argmin}_{\ell} \sum_j sim(\mathbf{u},\mathbf{v}_j) \cdot I(v_j, \ell)$ for each hidden variable $u$, it is immediate that the overall cost is minimized in Eq.~\eqref{eq:minsum} because the overall cost is defined as the sum of each hidden variable's cost.
\end{proof}

This theorem discusses a sufficient condition of the optimal min-sum solution but sometimes the sufficient condition, $\ell_u = \operatorname{argmin}_{\ell} \sum_j sim(\mathbf{u},\mathbf{v}_j) \cdot I(v_j, \ell)$ for each $u \in X$, is not achievable. However, what the min-sum does in such a case is to strategically drop the sufficient condition for some hidden variables to better minimize the sum of costs for other majority of hidden variables. Therefore, we can still say that the sufficient condition is achievable in general in any network for its majority of hidden variables. In particular, our embedding and hinge-loss based edge potential assignment bring large flexibility to the process. Therefore, the cost sum can be effectively minimized with the proposed method. Fig.~\ref{fig:emb} shows one such example that our proposed method is able to achieve the sufficient condition in most cases by ignoring some minor edges with high similarity. Because of this property, our approach is robust to evasion unless the attacker \emph{collectively evade} for neighboring URLs/domains/IP addresses/name servers (see Fig.~\ref{fig:evasion} for an example). However, the collective evasion will cost non-trivial expenses to the attacker.

\section{Experiments}\label{sec:Experiments}
In this section, we introduce our detailed experimental environments and results. We collected many URLs from crowd-sourced repositories and other papers. After that, we conducted experiments with ten baselines, ranging from classical classifiers and graphical methods to graph convolutional networks. Our method shows the best accuracy and robustness.

The source codes, data, and reproducibility information of our method are available at \url{https://github.com/taerikkk/BPE}.

\begin{table}[t]
\centering
\caption{The number of phishy and benign URLs for each dataset. Note that Sorio's and Ahmad's datasets are already tagged with ground-truth labels, so we did not use \url{virustotal.com} for them. There exist overlapped URLs so the total number of URLs is smaller than their sum.\label{tbl:data}}
\vspace{-0.5em}
{\renewcommand{\arraystretch}{1.3}
\begin{tabular}{|c|c|r|r|}
\hline
Dataset         & \begin{tabular}[c]{@{}c@{}}VirusTotal\vspace{-3px}\\Threshold\end{tabular} & \begin{tabular}[c]{@{}c@{}}\# Phishing\vspace{-3px}\\URL\end{tabular} & \begin{tabular}[c]{@{}c@{}}\# Benign\vspace{-3px}\\ URL\end{tabular} \\ \hline \hline
Bank of America & $4/7$                                                             & 4,610        & 9,408        \\ \hline
eBay            & $4/7$                                                             & 8,529        & 18,800       \\ \hline
PayPal          & $4/7$                                                             & 9,690        & 17,572       \\ \hline
Sorio et al.~\cite{Sorio2013DetectionOH} & N$/$A & 40,439 & 3,637 \\ \hline 
Ahmad et al.~\cite{ahmad} & N$/$A & 62,231 & 344,800 \\ \hline \hline
Total & N$/$A & 119,012 & 381,734 \\ \hline
\end{tabular}}
\end{table}

\subsection{Datasets}\label{data}
There have been created several phishing URL detection datasets~\cite{Ma09beyondblacklists,35580,Mohammad2014}. However, almost all of them do not release raw URL strings so we cannot use their datasets. We found only two open datasets with raw URL strings~\cite{Sorio2013DetectionOH,ahmad}. In addition to them, we also crawled \url{phishtank.com} and collected three sets of URLs reported during a couple of months recently for Bank of America, eBay, and PayPal, the top-3 most popular targets in the website (see Section~\ref{sec:crawl} for more details). \url{Phishtank.com} is a crowdsourced repository of suspicious URLs that does not provide ground-truth labels --- users can upvote or downvote the reported URLs in the website, but its voting system is not reliable because anyone (even including attackers) can participate. In total, we have about 500K URLs, 172K domains, and 66K IP addresses. Instead, we used \url{virustotal.com} to tag collected URLs. This website returns the prediction results of over 60 anti-virus (AV) products given a URL. The seven most reliable and popular AV products (such as McAfee, Norton, Kaspersky, Avast, and Trend Micro) were selected among them, and a URL is considered phishy if more than half of them indicate so, \textit{i.e.,} tagging by majority vote. At the end, we merged these datasets into one and created a very large URL dataset whose statistics are shown in Table~\ref{tbl:data}.

We split the combined set in the standard ratio of 80:20 for training and testing. Only 10\% of the URLs have timestamps. With them, we also tried to split in chronological order. Our method shows good accuracy for this configuration as well. However, we do not include the results because of i) its results similar to that of the random split, ii) its small data size, and iii) space reasons.

\begin{table*}[t]
\center
\caption{Detection results of some selected baseline methods and our proposed method. The best result in each measure (\textit{i.e.,} each column) is indicated in boldface. \textsc{\textsf{BPE}} is our method.}\label{tbl:result}
\vspace{-0.5em}
{{\renewcommand{\arraystretch}{1.2}
\begin{tabular}{|c|c|c|c|c|c|}
\hline
Type & Method &\begin{tabular}[c]{@{}c@{}}Recall (Phishy)\end{tabular} &\begin{tabular}[c]{@{}c@{}}Precision (Phishy)\end{tabular}& F-1 & Accuracy \\ \hline \hline
\multirow{6}{*}{Baseline}   & AdaBoost      & 0.830 & 0.830 & 0.830 & 0.831 \\ \cline{2-6}
                            & SGDClassifier & 0.762 & 0.720 & 0.734 & 0.720 \\ \cline{2-6}
                            & RandomForest  & 0.840 & 0.850 & 0.840 & 0.847 \\ \cline{2-6}
                            & LSTM          & 0.697 & 0.710 & 0.688 & 0.857 \\ \cline{2-6}
                            & 1DConv        & 0.677 & 0.735 & 0.689 & 0.864 \\ \cline{2-6}
                            & 1DConv+LSTM   & 0.788 & 0.806 & 0.784 & 0.902 \\ \hline \hline
\begin{tabular}[c]{@{}c@{}}Noisy\vspace{-3px}\\ Network\end{tabular} & \textsc{\textsf{BPE}} & 1.000 & 0.001 & 0.001 & 0.083 \\ \hline
\multirow{6}{*}{\begin{tabular}[c]{@{}c@{}}Simple\vspace{-3px}\\ Network\end{tabular}} & RWR & 0.569 &  0.917 & 0.702  & 0.815  \\ \cline{2-6} 
& POL & 0.901 & 0.853 & 0.876 & 0.943  \\ \cline{2-6} 
& \begin{tabular}[c]{@{}c@{}}\textsc{\textsf{BPE}}\vspace{-3px}\\(Cos, Deepwalk)\end{tabular} & 0.901 &  0.864 & 0.882 & 0.945 \\ \cline{2-6}
& \begin{tabular}[c]{@{}c@{}}\textsc{\textsf{BPE}}\vspace{-3px}\\(RBF, Doc2Vec)\end{tabular} & 0.895 & 0.864 &  0.879 &   0.943 \\ \hline
\multirow{8}{*}{\begin{tabular}[c]{@{}c@{}}Extended\vspace{-3px}\\ Network\end{tabular}} & RWR & 0.648 &  \textbf{0.930} & 0.764 & 0.863 \\ \cline{2-6} 
  & POL & 0.899  &  0.850   & 0.874  & 0.942 \\ \cline{2-6}
  & LGCN & \textbf{0.999}  &  0.762   & 0.865  & 0.762 \\ \cline{2-6} 
  & GAT & 0.995  &  0.762   & 0.863  & 0.760 \\ \cline{2-6} 
  & \begin{tabular}[c]{@{}c@{}}\textsc{\textsf{BPE}}\vspace{-3px}\\(Cos, Deepwalk)\end{tabular} & 0.958 & 0.831  & 0.890 & \textbf{0.969}\\ \cline{2-6}
 & \begin{tabular}[c]{@{}c@{}}\textsc{\textsf{BPE}}\vspace{-3px}\\(RBF, Deepwalk)\end{tabular} &  0.958  & 0.832 & \textbf{0.891}  & \textbf{0.969} \\ \hline
\end{tabular}}}
\end{table*}

\subsection{Baselines and Hyperparameters} Among many methods proposed, we consider the following baseline methods in our experiments. First, we test many feature-based prediction models. For this, we had surveyed literature and collected 19 features (see Appendix~\ref{sec:baseline}). After that, we predict with various classifiers after under/oversampling to address the imbalanced nature in our dataset --- benign URL numbers are much larger than phishing URL numbers in the training set.
In addition to synthetic minority oversampling~\cite{DBLP:journals/corr/abs-1106-1813} and adaptive synthetic sampling~\cite{He08adasyn:adaptive}, we consider five undersampling methods, six oversampling methods, and one ensemble method below.

Undersampling methods:
\begin{itemize}
    \item Naive random undersampling is randomly choose samples to drop.
    \item Tomek's link is a representative undersampling method.
    \item Clustering uses centroids of clusters after dropping other cluster members.
    \item NearMiss is also popular for undersampling.
    \item Various nearest neighbor methods are able to undersampling.
\end{itemize}

Oversampling methods:
\begin{itemize}
    \item Naive random oversampling is randomly choose samples to add.
    \item SMOTE~\cite{DBLP:journals/corr/abs-1106-1813} and its variants are a family of the most popular oversampling methods, which include five variations.
    \item ADASYN~\cite{He08adasyn:adaptive} is also popular for oversampling.
\end{itemize}

Ensemble method:
\begin{itemize}
    \item Ensemble method means that we use both the oversampling and undersampling methods at the same time.
\end{itemize}

We refer to a survey paper~\cite{JMLR:v18:16-365} for more detailed information.
The combination of classifiers, under/oversampling methods, and their hyperparameters create a huge number of possible options in this method. So, first, we perform 5-fold cross validation to choose the best performing classifier/sampling method, and its hyperparameters. Second, we also test three deep learning-based sequence classification methods mentioned in Section~\ref{sec:rel}. These neural networks are based on recurrent or convolutional layers. We use their hyperparameters recommended in their original publications. 

Third, on the \textit{simple network} that consists only of URLs and their words, we run the following graphical methods: i) Random Walk with Restart (RWR): This method runs many random walks from training URLs and counts the number of visits to each testing URL. It is very successful for recommender systems~\cite{Cooper:2014:RWR:2567948.2579244}. ii) Polonium (POL): Polonium based on a simple belief propagation strategy showed a big success in predicting malware and malicious domains. We run the belief propagation on our network with Polonium's compatibility matrix definition in Table~\ref{table:polonium}. iii) Belief Propagation with Enhancements (\textsc{\textsf{BPE}}): This is our method to run the belief propagation based on our improved definition of compatibility matrix. We test various embedding techniques, $ths_+ = \{0, 0.1, 0.3, 0.5, 0.7, 0.9, 1\}$, $ths_- = \{0, 0.1, 0.3, 0.5, 0.7, 0.9, 1\}$, and for calculating the vector similarity, the cosine similarity and RBF kernel. We set the dimension of the embeddings to 128.

Fourth, on the \textit{extended network} that consists of all entities (cf. Section~\ref{sec:ne}), we test the same set of graphical methods: RWR, POL, and \textsc{\textsf{BPE}}. For this, we use the blacklist of 41,881 IP addresses and 158,271 domains provided by \url{virustotal.com}. In other words, those blacklisted entities are converted into observed variables and excluded from the inference process. We also test \textsc{\textsf{BPE}} on the \textit{noisy network} where stop words are not removed. Last, we test state-of-the-art graph convolutional networks (GCNs) such as LGCN~\cite{Gao:2018:LLG:3219819.3219947} and GAT~\cite{velickovic2018graph} on the extended network. For a vertex $v$, we feed a feature vector after concatenating i) the 19 features of $v$ we use in the feature-based prediction, ii) a binary value denoting whether $v$ is blacklisted or not, and iii) a one-hot vector where only the index of the vertex $v$ is one. If some items are missing, we concatenate with zeros --- \textit{e.g.,} a domain does not have the 19 features so we zero them out. We test the hyperparameters recommended in their original papers. To prevent overfitting, we also add an L2 regularization of neural network weights. In all those graphical models, such as RWR, POL, GCNs, and our method (\textit{i.e.,} \textsc{\textsf{BPE}}), the labels of training URLs are fixed and only unknown labels of testing URLs are inferred.

We exclude other content-based detection methods in our experiments because it is hard to obtain web page contents in general --- recall that phishing attacks do not last long and attackers usually clean their traits from the Internet after the accomplishment of their goal. The two datasets we downloaded from ~\cite{Sorio2013DetectionOH,ahmad} do not include any content information and we also could not collect web page information in HTML from \url{phishtank.com} in a stable manner.

\subsection{Environments}
\paragraph{Hardware} We conducted our experiments on the machines with i9-9900K, 64GB RAM, and GTX 1070.
\paragraph{Software} As our experiments utilize many different types of baseline methods, our software environments are rather complicated. The selected list of important software/libraries are as follows:
\begin{itemize}
    \item Python ver 3.8.1.
    \item Scikit Learn ver 0.22.1.
    \item TensorFlow ver 1.5.1.
    \item CUDA ver 10.
    \item NetworkX ver 2.4.
\end{itemize}

\vspace{-0.5em}
\subsection{Experimental Results} We summarize the results shown in Table~\ref{tbl:result} as follows. Among all feature-based methods, RandomForest performs the best. For all metrics, it outperforms AdaBoost, SGDClassifier, and others, \textit{e.g.,} the F-1 score of 0.840 for RandomForest vs. 0.830 for AdaBoost vs. 0.734 for SGDClassifier. However, all these feature-based baseline methods are clearly beaten by the network-based methods. This supports the efficacy of our network-based approach.

RWR's precision for the phishy class on the extended network is the best (0.930). However, its recall is worse than other network-based inference methods. POL shows a balanced performance between recall and precision as in its original task to detect malware. LGCN's recall for the phishy class is the best (0.999). For LGCN and GAT, we found that they are sensitive to hyperparameters and hard to regularize the overfitting. Surprisingly, the best F-1 was made when we allow overfitting to the phishy class to some degree. When we increase the coefficient of the L2 regularizer to prevent overfitting, their F-1 scores drastically decrease. We also found that training with subgraphs is not effective in processing our large network.
Therefore, we set the size of subgraph as large as possible in our recent GPU model --- due to the GPU memory limitation, whole graph training is impossible for our network --- but its performance is inferior to our method.

Our method with $ths_+ = 0.7$, $ths_- = 0.7$, RBF kernel, and DeepWalk, which is marked as `\textsc{\textsf{BPE}} (RBF, Deepwalk)', shows the best performance for F-1 and accuracy. Although \textsc{\textsf{BPE}}'s precision for the phishy class is a little lower (0.832) than the best feature-based method (\textit{i.e.,} RandomForest)'s score (0.850), the \textsc{\textsf{BPE}}'s recall for the phishy class is much higher (0.958) than that of RandomForest (0.840).
However, one may be worried that our method mis-classifies benign as phishy due to its relatively low precision.
To this end, we measure the false positive rate (\textit{i.e.,} FPR) to \textsc{\textsf{BPE}} and RandomForest.
As a result, we could obtain 0.031 for \textsc{\textsf{BPE}} and 0.306 for RandomForest. Therefore, we expect that \textsc{\textsf{BPE}} is the most useful for accurately detecting phishing URLs in practice.

The same network-based method on the noisy network shows poor performance (\textit{e.g.,} 0.01 for F-1), which proves our network definition also plays an important role.

\begin{table*}
\centering
\caption{F-1 scores of \textsc{\textsf{BPE}}, POL and RandomForest (RF) after M1-5 evasions. The best result in each evasion method and ratio is indicated in boldface. \textsc{\textsf{BPE}} is our method.}\label{tbl:evasion}
\vspace{-0.5em}
{\renewcommand{\arraystretch}{1.2}
\begin{tabular}{|c|c|c|c|c|c|c|c|c|c|c|c|c|c|c|c|}
\hline
\multirow{2}{*}{\begin{tabular}[c]{@{}c@{}}Evasion\\Ratio\end{tabular}} & \multicolumn{3}{c|}{M1 (domain)} & \multicolumn{3}{c|}{M2 (path)} & \multicolumn{3}{c|}{M3 (query)} & \multicolumn{3}{c|}{M4 (domain and path)} & \multicolumn{3}{c|}{M5 (domain and query)} \\ \cline{2-16} 
 & \textsc{\textsf{BPE}} & POL & RF & \textsc{\textsf{BPE}} & POL & RF & \textsc{\textsf{BPE}} & POL & RF & \textsc{\textsf{BPE}} & POL & RF & \textsc{\textsf{BPE}} & POL & RF \\ \hline \hline
5\% & \textbf{0.866} & 0.836 & 0.812 & \textbf{0.876} & 0.843 & 0.820 & \textbf{0.888} & 0.867 & 0.821 & \textbf{0.861} & 0.811 & 0.813 & \textbf{0.873} & 0.822 & 0.814 \\ \hline
10\% & \textbf{0.847} & 0.817 & 0.816 & \textbf{0.861} & 0.822 & 0.810 & \textbf{0.882} & 0.841 & 0.816 &  \textbf{0.829} &  0.778 & 0.804 &  \textbf{0.863} &  0.811 & 0.807 \\ \hline
15\% & \textbf{0.833} & 0.802 & 0.810 & \textbf{0.858} & 0.817 &  0.803 & \textbf{0.882} & 0.836 & 0.811 &  \textbf{0.805} &  0.760 & 0.790 &  \textbf{0.854} &  0.807 & 0.798 \\ \hline
\end{tabular}}
\end{table*}
\vspace{-0.8em}

\begin{table}
\centering
\caption{F-1 scores of \textsc{\textsf{BPE}}, POL and RandomForest (RF) after M6 and M7 evasions. The best result in each evasion method and ratio is indicated in boldface. \textsc{\textsf{BPE}} is our method.}\label{tbl:evasion2}
\vspace{-0.5em}
{\renewcommand{\arraystretch}{1.2}
\begin{tabular}{|c|c|c|c|c|c|c|}
\hline
\multirow{2}{*}{\begin{tabular}[c]{@{}c@{}}Evasion\\Ratio\end{tabular}} & \multicolumn{3}{c|}{M6 (path and query)} & \multicolumn{3}{c|}{M7 (all)} \\ \cline{2-7} 
 & \textsc{\textsf{BPE}} & POL & RF & \textsc{\textsf{BPE}} & POL & RF \\ \hline \hline
5\% & \textbf{0.874} & 0.838 & 0.814 & \textbf{0.861} & 0.827 & 0.760 \\ \hline
10\% & \textbf{0.869} & 0.832 & 0.808 & \textbf{0.828} & 0.791 & 0.751 \\ \hline
15\% & \textbf{0.857} & 0.820 & 0.802 & \textbf{0.803} & 0.762 & 0.733 \\ \hline
\end{tabular}}
\end{table}

\paragraph{Statistical significance} For the statistical significance of our experiments, we conduct paired $t$-tests with a 95\% confidence level between \textsc{\textsf{BPE}} and each baseline, and achieved a $p$-value less than 0.05 for all cases.

\paragraph{Transductive vs. Inductive}
Transductive and inductive inferences are two popular paradigms of machine learning~\cite{Vapnik:1995:NSL:211359}. Among all the baseline methods, RandomForest and some other classifiers are inductive methods and many other network-based methods are transductive. In many cases, people rely on the inductive inference where a generalized prediction model trained with a training set predicts for unknown testing samples. In our work, however, we adopt a transductive method where the class label of a specific unknown testing sample is inferred from other specific related training samples in the network architecture. Fig.~\ref{fig:network} justifies our transductive approach because a cluster usually consists of vertices from the same class. However, it is not the case that all transductive methods are successful in Table~\ref{tbl:result}.

\paragraph{Time performance}
\textsc{\textsf{BPE}} is an advanced LBP-based method with our novel similarity-based edge potential assignments. Therefore, the time complexity of \textsc{\textsf{BPE}} is $O(S \cdot |E| \cdot t)$, where $S$ indicates a similarity calculation cost, $E$ indicates the set of edges, and $t$ does the number of iterations required for the convergence. $t$ is typically small in our setting, \textit{e.g.,} $t=5$ is enough. The time complexity of RandomForest (\textit{i.e.,} the best feature-based method) is $O(f \cdot n \cdot \log(n))$, where $f$ is the number of features and $n$ is the number of URLs. In our experiments, the training (wall-clock) time of \textsc{\textsf{BPE}} is 4.7 times faster than that of RandomForest.

\subsection{Parameter Sensitivity}
\paragraph{Sensitivity to thresholds} The following threshold combinations perform very well and are comparable to each other in our experiments: ($ths_+$ = 0.7, $ths_-$ = 0.7), ($ths_+$ = 0.3, $ths_-$ = 0.9), ($ths_+$ = 0.3, $ths_-$ = 0.5), ($ths_+$ = 0.7, $ths_-$ = 0.9), ($ths_+$ = 0.5, $ths_-$ = 0.3), and so on. One common characteristic of them is that two extreme values, 0 and 1, are not preferred. This supports our decision to adopt thresholds because two dissimilar neighbors do not always mean that their labels should be different. In other words, the one you are not close to is not necessarily your enemy. By limiting the penalty, we achieved the best accuracy in our experiments.

\paragraph{Sensitivity to embedding} It turns out that network embeddings are more effective than word or document embedding methods. All high ranked results are produced by DeepWalk. Doc2Vec produces the best result only for the simple network and RBF kernel environment. We think that this is because our network definition considers common words among URLs and DeepWalk is able to capture the semantic meanings of words closely located in the network.

\paragraph{Cosine similarity vs. RBF kernel} It seems the cosine similarity  and the RBF kernel are comparable to each other in our experiments. When sorting all results, all highly ranked results are evenly distributed to both of them.

\begin{figure*}[t]
    \centering
    \subfloat[Original]{\includegraphics[width=0.248\textwidth]{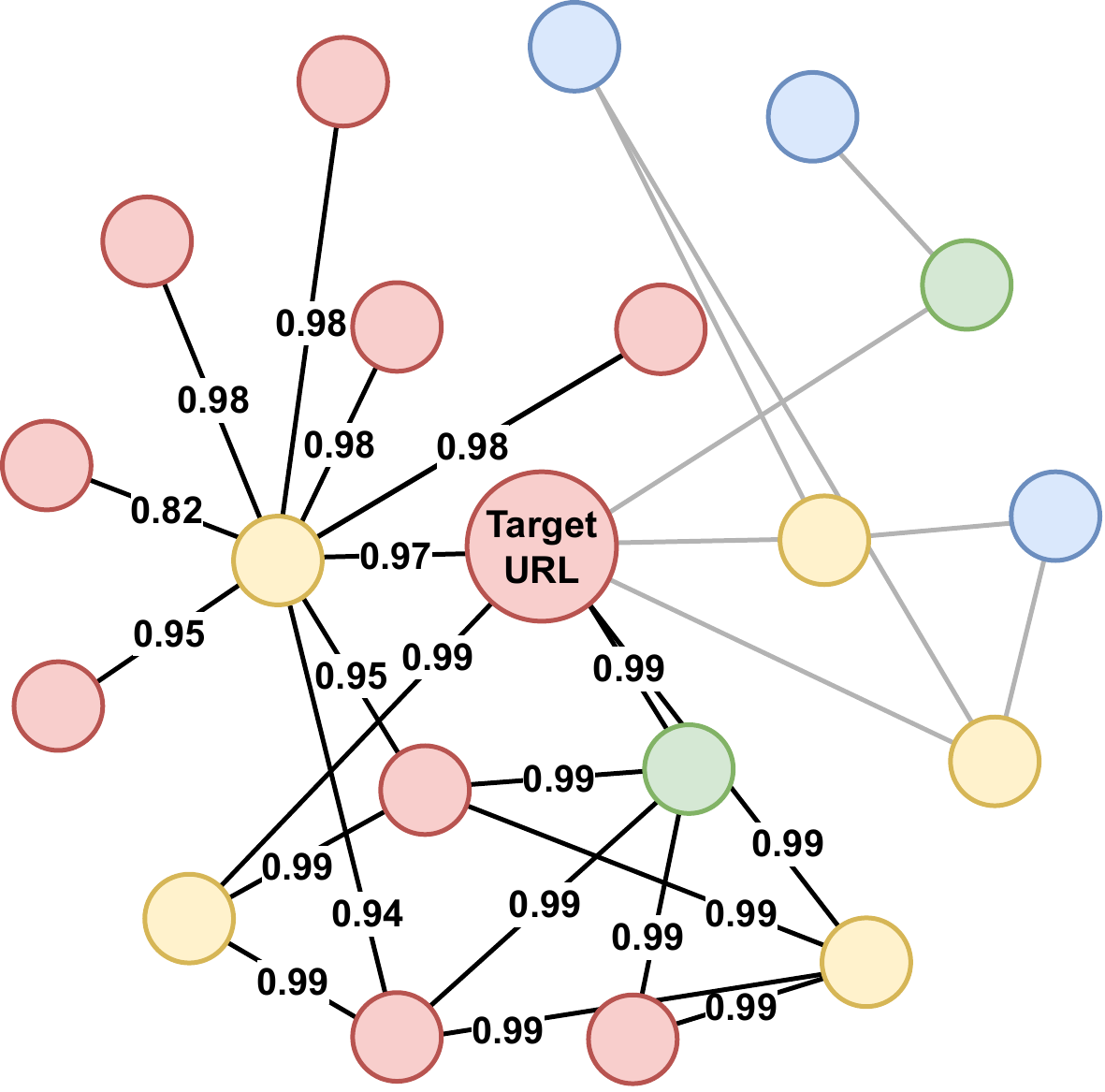}}\hfill
    \subfloat[M1 Evasion]{\includegraphics[width=0.248\textwidth]{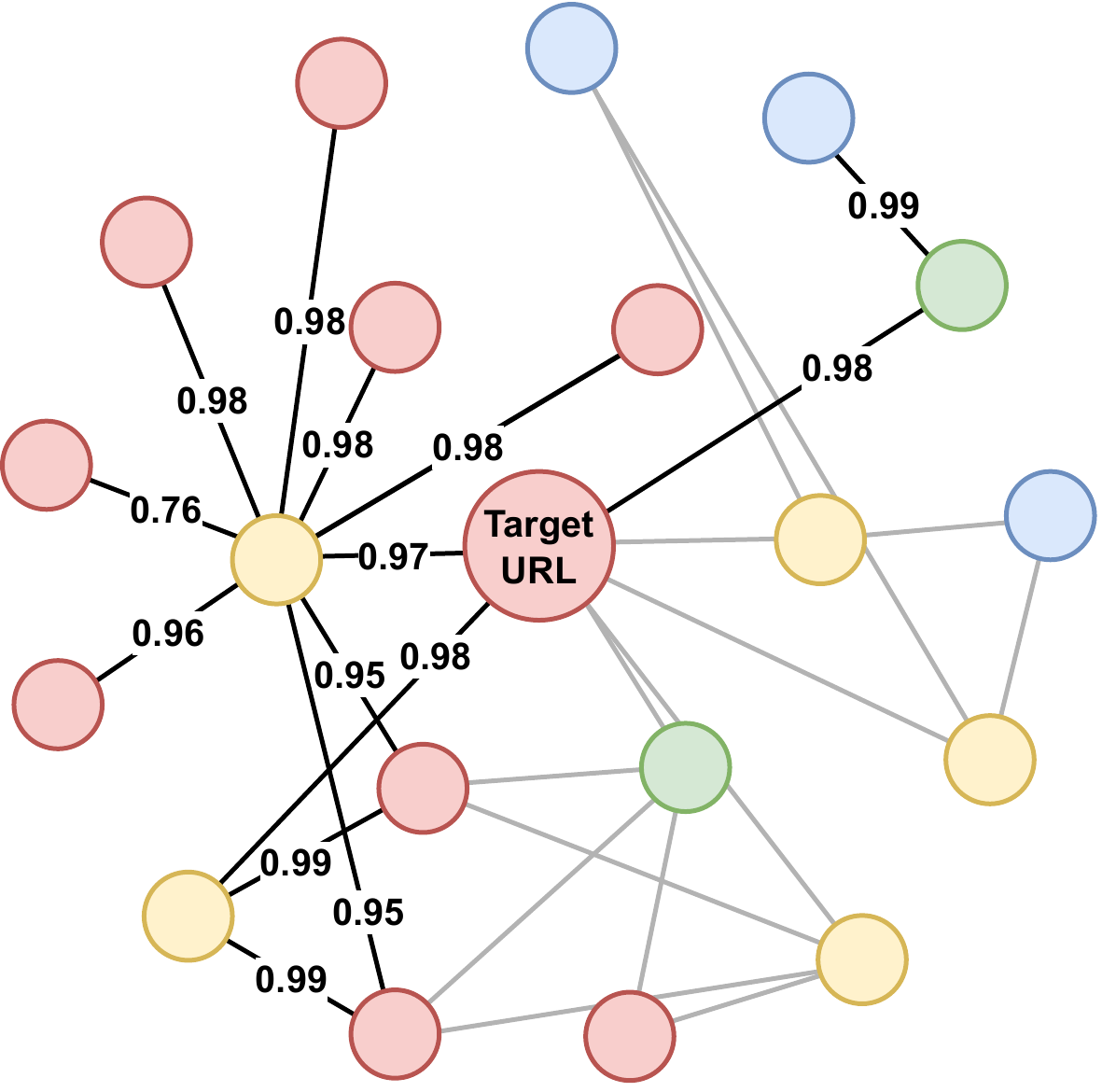}}\hfill
    \subfloat[M2 Evasion]{\includegraphics[width=0.248\textwidth]{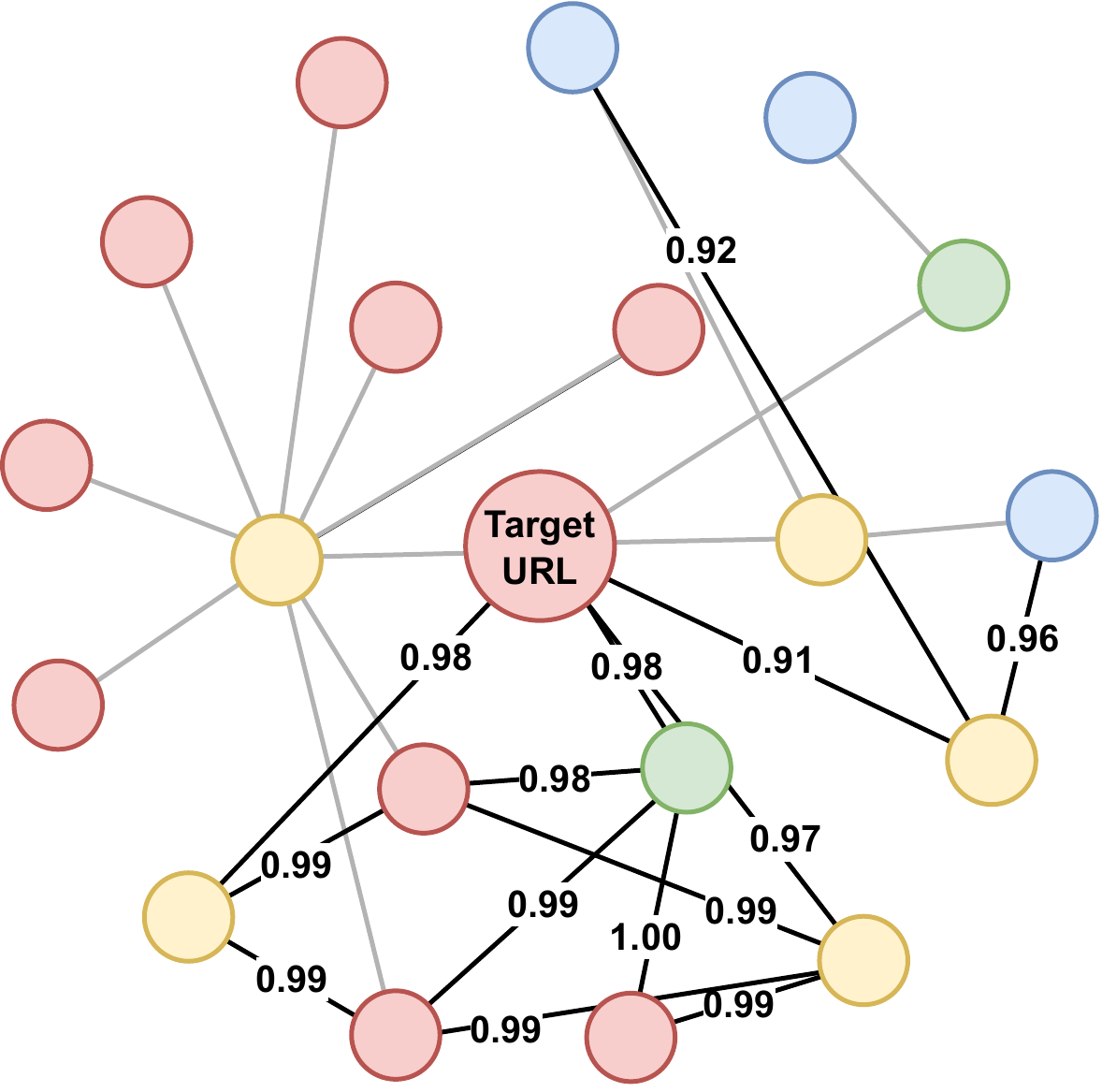}}\hfill
    \subfloat[M3 Evasion]{\includegraphics[width=0.248\textwidth]{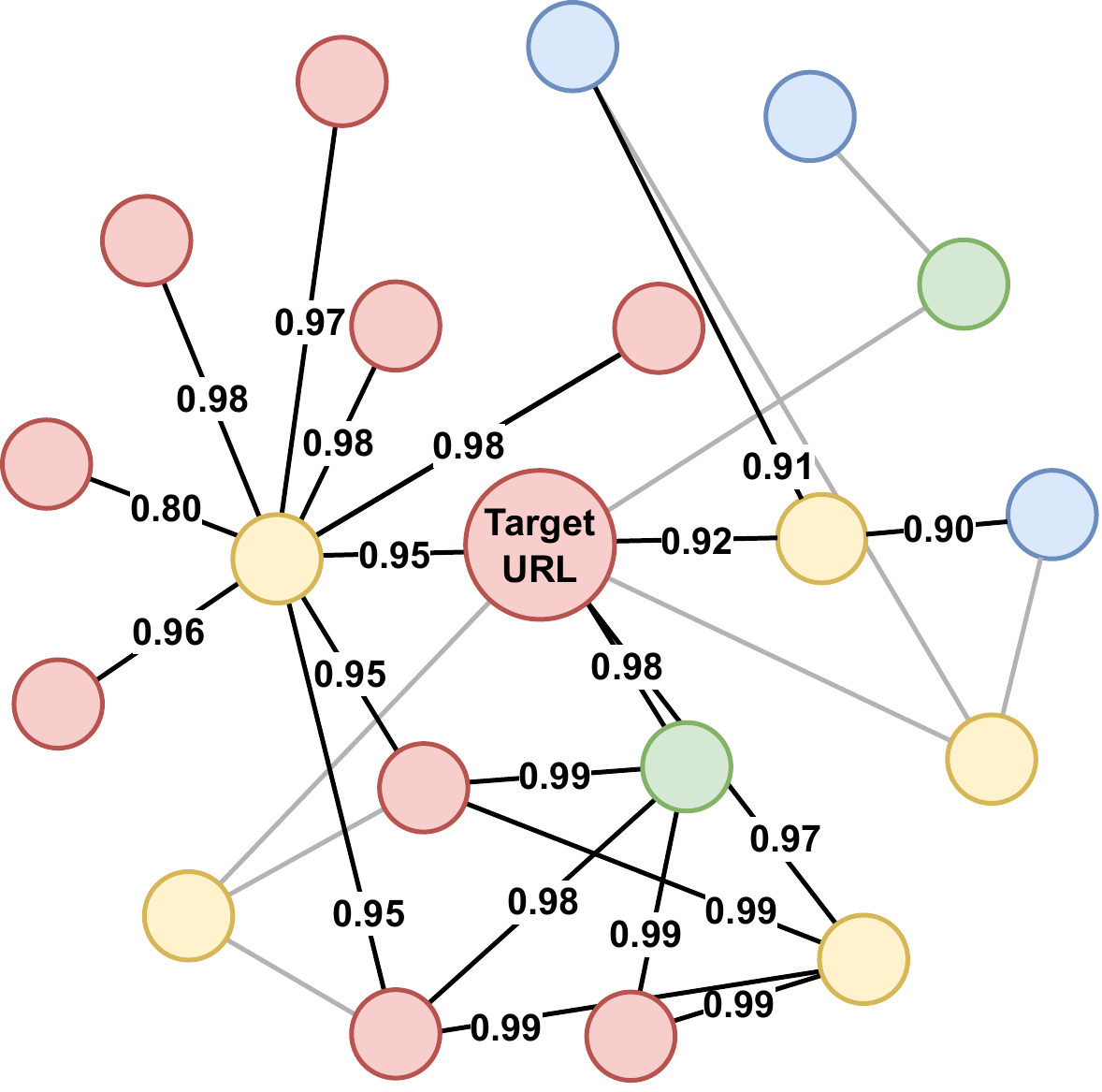}}\hfill
    \subfloat[M4 Evasion]{\includegraphics[width=0.248\textwidth]{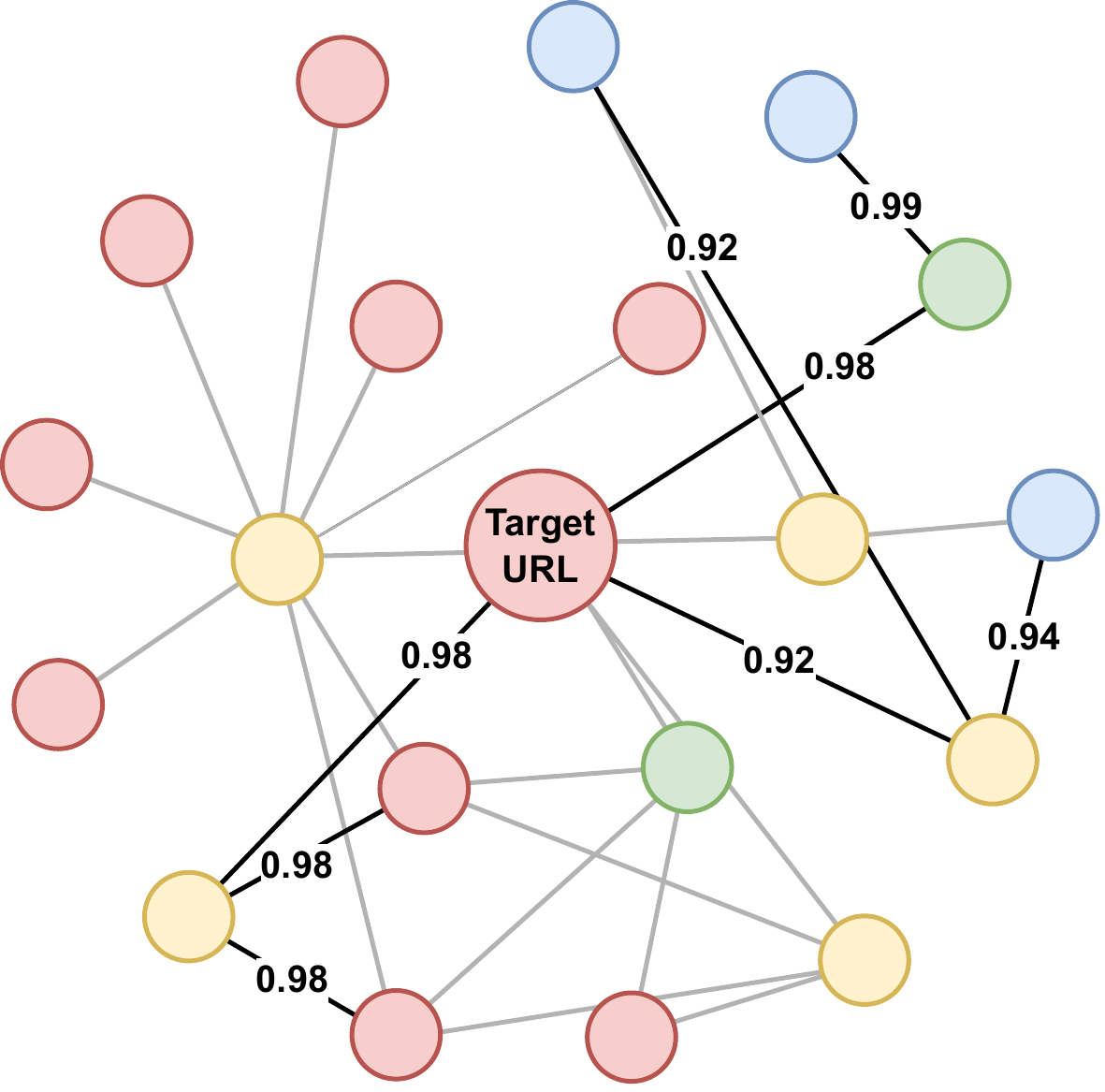}}\hfill
    \subfloat[M5 Evasion]{\includegraphics[width=0.248\textwidth]{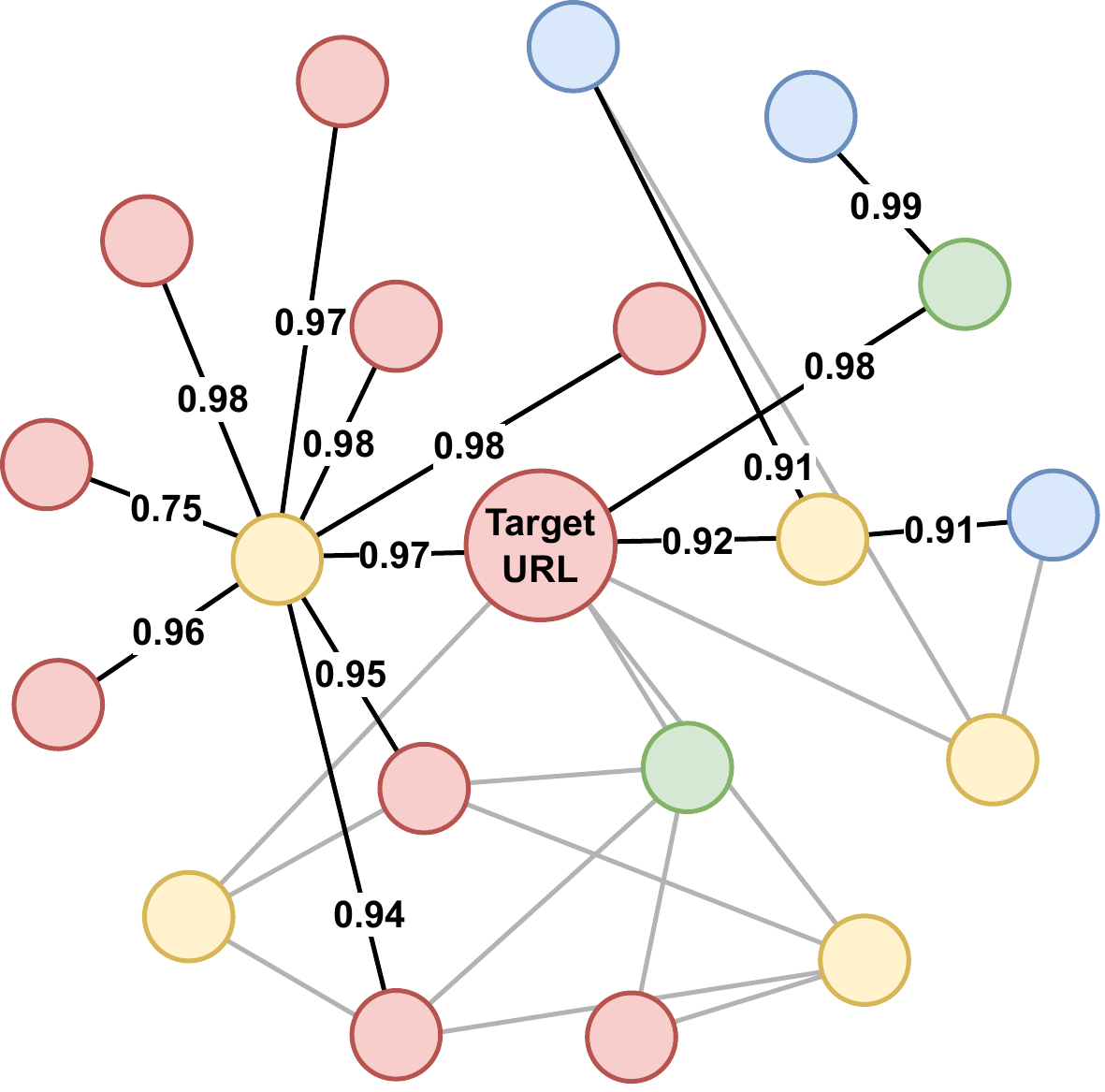}}\hfill
    \subfloat[M6 Evasion]{\includegraphics[width=0.248\textwidth]{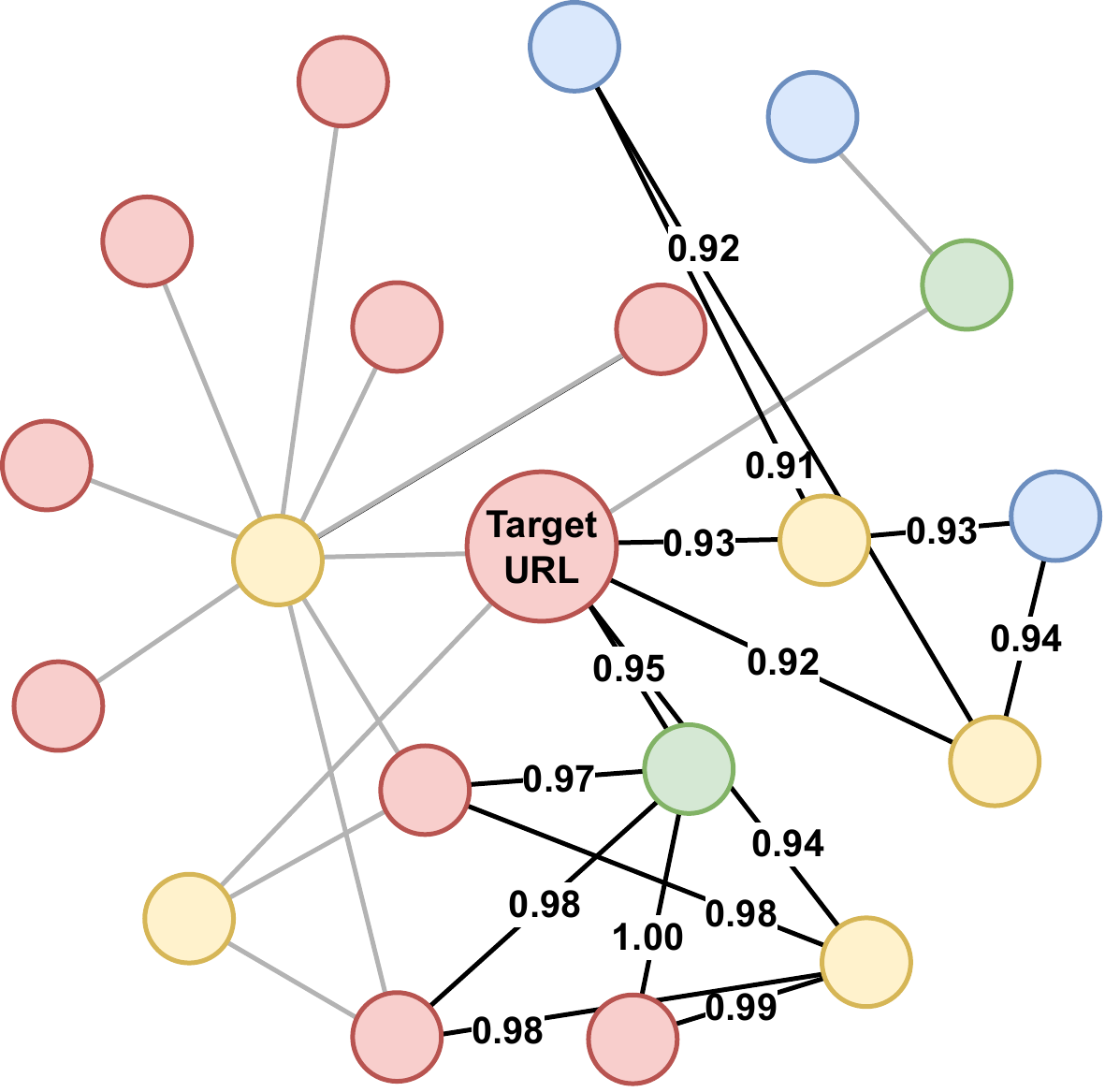}}\hfill
    \subfloat[M7 Evasion]{\includegraphics[width=0.248\textwidth]{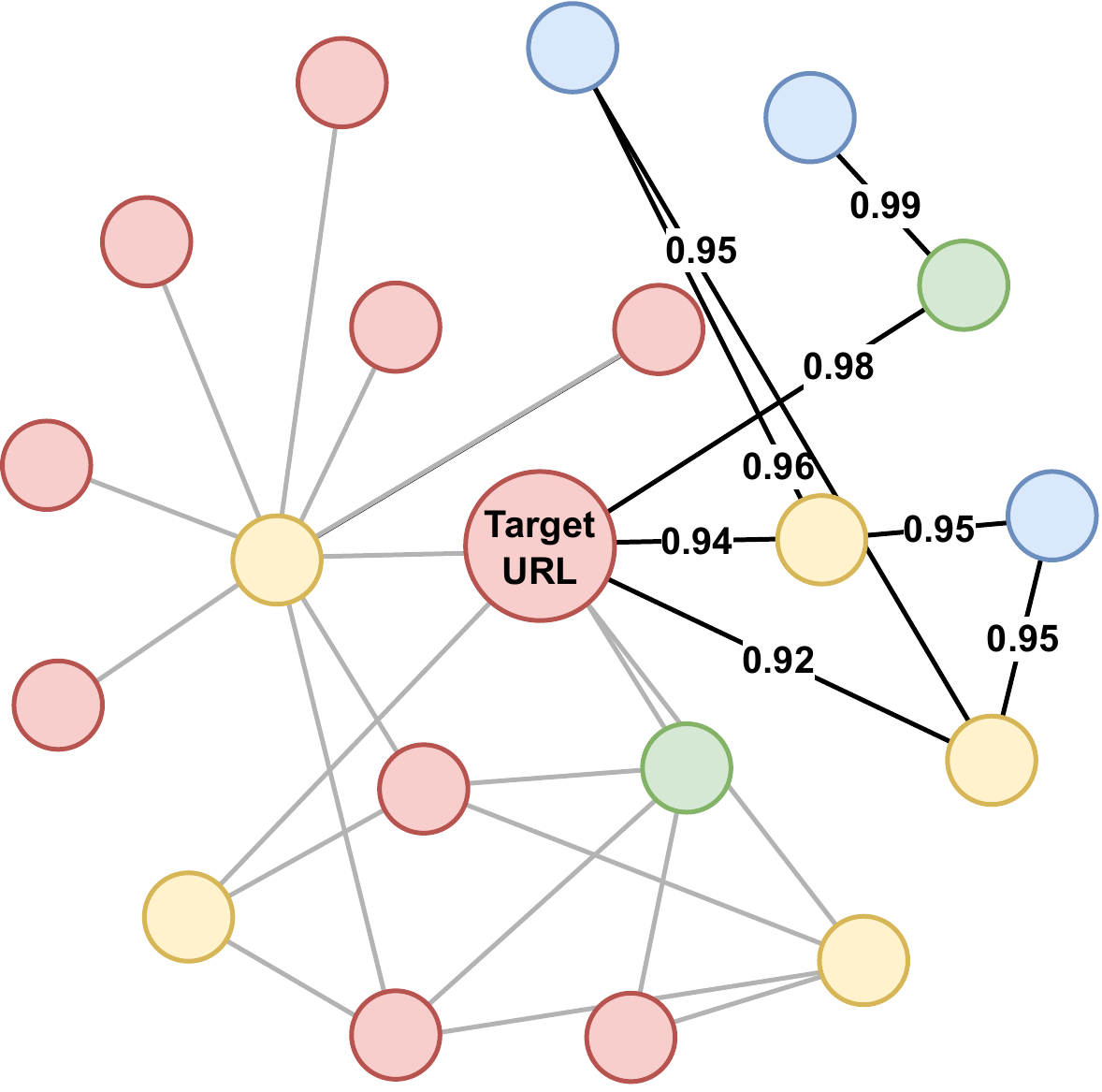}}\hfill
    \caption{Visualization of the original network and  M1-7 evasions of the target phishing URL denoted with the largest red vertex. Each edge is annotated with the similarity. The meaning of vertex color follows that in Fig.~\ref{fig:network}.}
    \label{fig:evasion_case}
\end{figure*}

\subsection{Evasion Tests}\label{sec:evasion}
For our evasion testing, we consider all possible variations for the parts of phishing URLs, \textit{i.e.,} domain, path, and query. Specifically, we define, in total, seven evasion methods (\textit{i.e.,} M1-7) as follows: M1) Phishing URL's domain is changed to other random benign domain (and as a result, IP address is changed too); M2) Phishing URL's path string (cf. Section~\ref{sec:seg}) is changed to other random benign one; M3) Phishing URL's query string (cf. Section~\ref{sec:seg}) is changed to other random benign one; M4) Phishing URL's domain and path string are changed to other random benign ones; M5) Phishing URL's domain and query string are changed to other random benign ones; M6) Phishing URL's path and query strings are changed to other random benign ones; M7) Phishing URL's each part is independently changed to other random benign ones, \textit{i.e.,} phishing URL becomes an entirely new URL that looks like benign.

Note that our evasion tests embrace Shirazi et al.'s evasion settings (cf. Section~\ref{sec:eva}). Also, we note that M7 evasion is the most challenging situation. As mentioned earlier, for M7 evasion, the attackers' motivation may be low, because it requires non-trivial expenses. Nevertheless, it is worth mentioning that we take into account the case where all of the domain, path, and query strings are evaded simultaneously.

For some spear phishing attacks aiming at particular targets, however, sophisticated URLs are prepared with all benign string patterns and web page contents, in which case more advanced techniques are required to detect. It is well known that the attacker invests large efforts in spear phishing by considering even the psychological and habitual characteristics of the targets after hijacking benign user accounts~\cite{Oliveira:2017:DSP:3025453.3025831,Lin:2019:SSE:3349608.3336141,Ho:2019:DCL:3361338.3361427}. However, it is out of the scope of this paper and we leave it as our future work.

To simulate evasions, we modify random 5-15\% of our testing phishing URLs using one of the seven evasion methods. After the modifications, its network is also reconstructed accordingly.
We compare \textsc{\textsf{BPE}} (our method) with POL and RandomForest (RF) which represent network-based and best feature-based baseline methods, respectively. Because all entities are connected in our network and one evasion may affect other neighboring non-evaded URLs in the worst case, a simple measure counting the number of successful detections for the phishing URLs with evasion is not a correct metric. So, we re-evaluate all testing URLs again after evasion and report the results in Tables~\ref{tbl:evasion} and ~\ref{tbl:evasion2}.

As shown in Tables~\ref{tbl:evasion} and ~\ref{tbl:evasion2}, our \textsc{\textsf{BPE}} outperforms other baselines with non-trivial margins. Especially, \textsc{\textsf{BPE}} outperforms RandomForest by up to 13.29\% in the most challenging situation, \textit{i.e.,} M7 with an evasion ratio of 15\%.
In addition, although M7 evasion is the most challenging situation, where we independently change every part of a phishing URL to benign, \textsc{\textsf{BPE}} still shows relatively high F-1 scores (0.803-0.861).
This is because each part of a benign URL connected to a phishing URL is not likely to have high similarity scores (since these are from different benign URLs), so the phishing URL has a low similarity to each newly connected vertex.
Therefore, \textsc{\textsf{BPE}} with our novel similarity-based edge potential will not predict this phishing URL as benign.
On the other hand, POL using a majority voting of neighbors shows low F-1 scores (0.762-0.827) in M7 evasion.

Furthermore, we found that \textsc{\textsf{BPE}} in various evasion settings outperforms most baselines in non-evasion settings. Specifically, except for M1 with an evasion ratio of 15\% and M4 (resp. M7) with an evasion ratio of 10\% and 15\%, the minimum F-1 score of \textsc{\textsf{BPE}} in various evasion settings is 0.847 (\textit{i.e.,} M1 with an evasion ratio of 10\%), which surpasses that of the best baseline in non-evasion settings, \textit{i.e.,} 0.840 for RandomForest.
One more important fact is that evasion incurs additional costs to the attacker. To make a domain whitelisted, for instance, the attacker should pay hosting fees and maintain the domain for a considerable amount of time without any attack campaigns or should compromise other benign web servers. Some attackers do this and switch to phishing web pages at D-Day to launch a phishing attack~\cite{apwg}. Even after the attacker's efforts, experiments show that our method is good at detecting those evasion cases.

\begin{figure*}
    \centering
    \subfloat[Ground-truth]{\includegraphics[width=0.31\textwidth]{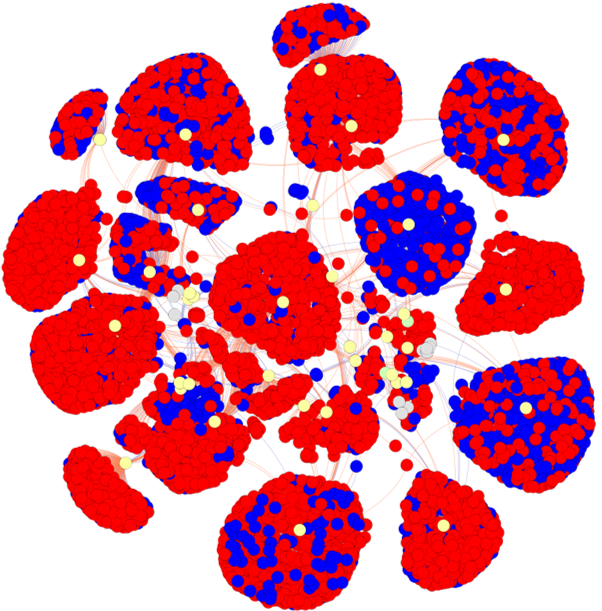}}\hfill
    \subfloat[Proposed Method, \textit{i.e.,} \textsc{\textsf{BPE}}]{\includegraphics[width=0.31\textwidth]{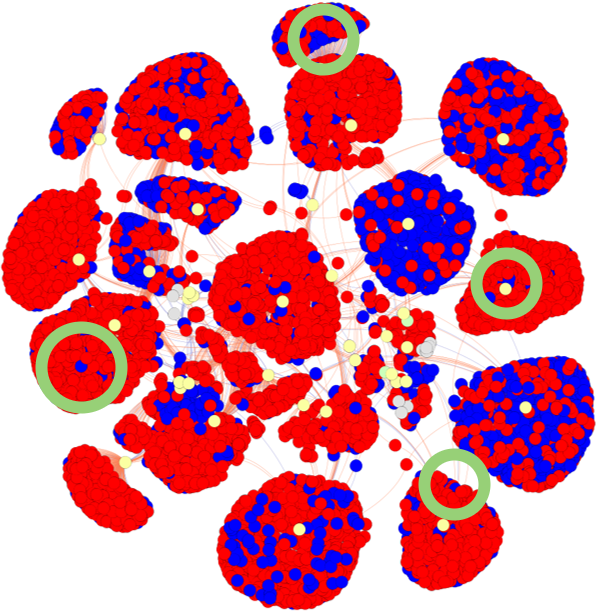}}\hfill
    \subfloat[RandomForest]{\includegraphics[width=0.31\textwidth]{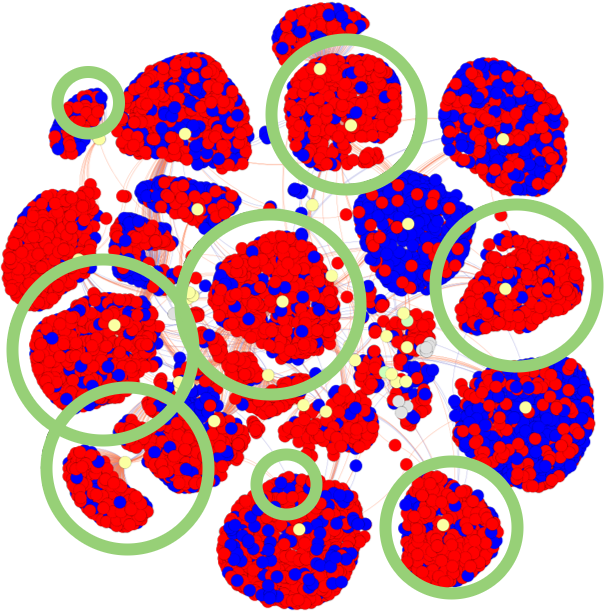}}
    \caption{Visualization of phishy/benign predictions for a partial area in our network. RandomForest, the best inductive method in our experiments, do not use our network information so, when being projected onto it, its predictions do not strictly follow the network connectivity as shown in (c). The meaning of vertex color follows that in Fig.~\ref{fig:network}.}\label{fig:trans}
    \vspace{-1em}
\end{figure*}

\paragraph{Evasion case study}
Fig.~\ref{fig:evasion_case} shows eight 2-hop ego networks for a phishing URL that is randomly selected for our evasion settings. The first one shows the original network connection in our dataset. The target URL (the largest red vertex) is connected to other phishy domain and words, in which case it is straightforward to classify the target URL as phishy. In the other seven networks, however, the target URL is connected to a benign domain or/and word(s). Even after these evasions, our method correctly infers that the target URL is still phishy whereas POL and RandomForest fail to detect all the evasion cases. Our method is equipped with a sophisticated edge potential assignment mechanism whereas POL does not consider them. Our theoretical analyses in Section~\ref{sec:rob} also well supports the robust nature of our method.

We also introduce other visualizations with real prediction results. Fig.~\ref{fig:trans} shows three visualizations including our method's and RandomForest's predictions. To emphasize their differences, we choose some important domain/IP/word vertices from our network and show their URL neighbors (rather than showing the full network). In Fig.~\ref{fig:trans} (a), we can observe a strong pattern that the ground-truth label follows the network connectivity in many cases. Sometimes red (phishy) and blue (benign) vertices are mixed in a cluster but this is mainly because we find the clusters in the sub-network only. Our method in Fig.~\ref{fig:trans} (b) shows a better compliance to the network connectivity than that in Fig.~\ref{fig:trans} (c). To evade our method, therefore, the majority of URLs in the same cluster should be evaded at the same time, which burdens the attacker with non-trivial costs (see our evasion cost discussion in Section~\ref{sec:eva}).

\section{Data Crawling}\label{sec:crawl}
To collect as many phishing URL samples as possible, we had monitored \url{phishtank.com} for a couple of months while searching other researchers' available data. There are several online datasets --- many of them were released by Ma et al. who had published several papers for phishing URL detection~\cite{Ma:2009:ISU:1553374.1553462,Ma09beyondblacklists}. However, their data does not include raw string patterns. We also contacted them but they replied that they cannot share the raw data. Mohammad et al. also released their data in \url{https://archive.ics.uci.edu/ml/datasets/Phishing+Websites} but they also do not release their raw data used for their research~\cite{DBLP:conf/icitst/MohammadTM12,Mohammad2014}. As mentioned earlier, we need string patterns of phishing URLs so we couldn't utilize all the above mentioned data.

Therefore, we programmed a web crawler using an automated web browser library and collected all the URLs reported for Bank of America, eBay, and PayPal. For retrieving additional information from \url{virustotal.com}, we received an academic license to their APIs and collected many such information we listed in the main paper. The academic license was activated for three months so it was more than enough for us to retrieve all the needed information.

\section{Conclusions \& Future Work}\label{sec:conclusions}
Although many (machine learning) methods have been proposed to detect phishing URLs, it had been overlooked that attackers can use evasion techniques to neutralize them. In this paper, we tackled the significant problem of detecting phishing URLs after evasion. After segmenting URLs into words and creating a heterogeneous network that consists of cross-related entities, we performed the belief propagation equipped with our customized edge potential mechanism which is our main contribution. Furthermore, we showed that our design is theoretically robust to evasion. We collected recent URLs and downloaded other two datasets for extensive experiments. Our experiments with about 500K URLs verify that our method is the most effective in detecting phishing URLs and also is the most robust to evasion than all baselines. Besides, we expect that our method can be easily applied to address any similar network-based problem (\textit{e.g.,} detecting fake accounts in social networks and email spam) if it can be represented as a classification on graphs.

In the future, we will study a string and content-based robust detection method. For some evasion techniques, it is limited to only string-based detection methods. However, it requires non-trivial efforts to collect web page contents. Therefore, we think that hybrid methods will be the most useful for real-world applications.

\section*{Acknowledgment}
The work of Sang-Wook Kim was supported by the Institute of Information \& Communications Technology Planning \& Evaluation (IITP) grant funded by the Korea government (MSIT) (No. RS-2022-00155586) and by the National Research Foundation of Korea (NRF) grant funded by the Korea government (MSIT) (No. 2018R1A5A7059549). The work of Noseong Park was supported by the Institute of Information \& Communications Technology Planning \& Evaluation (IITP) grant funded by the Korean government (MSIT) (No. 2020-0-01361, Artificial Intelligence Graduate School Program (Yonsei University)).

\bibliographystyle{ACM-Reference-Format}
\bibliography{CCS22}

\begin{appendices}
\section{Baseline Lexical, Host, and Domain Features}\label{sec:baseline}
We did an extensive literature survey and collected 19 features from the papers mentioned in our related work section. The complete list of the features we used in our experiments (sorted by the feature importance extracted from RandomForest, the best performing feature-based classifier in our experiments) is as follows:

\begin{enumerate}[itemindent=4em, label={(Ranking \raisenth*)}]
  \item \emph{Kullback-Leibler (KL) divergence}
  \newline The Kullback-Leibler (KL) divergence is a popular metric to measure the similarity between two probability distributions. We can calculate the KL divergence on character distributions between a URL and  the English language. The reference for the character distribution in the English language is obtained from ~\cite{characterFreq}.
  
  \item \emph{Entropy of URL}
  \newline It was found that the string entropy of a URL is an important feature, since many phishing URLs have random text, causing their entropies to be higher than those of benign URLs.
  
  \item \emph{Digit/Letter Ratio in the whole URL}
  \newline The ratio of digits w.r.t letters in the whole URL is also important.

  \item \emph{Top-level domain numbers in path}
  \newline Attackers often try to impersonate legitimate websites by adding multiple top-level domains in the path of a URL. If the count of top-level domains in the path exceeds one, then it is likely to be phishy.
  
  \item \emph{The number of dashes in path}
  \newline This is to count the occurrence of `-' in the path. Many dashes indicate phishing URLs.
  
  \item \emph{Blacklist}
  \newline A blacklist contains a set of malicious domains and IP addresses. If a URL has such a domain or an IP address, it can be immediately predicted as phishy. However, it is often incomplete and there are many missing malicious domains and IP addresses. In general, we do not use a whitelist because  attackers sometimes compromise whitelisted servers to implant their phishing web pages and we cannot always trust whitelisted ones.
  
  \item \emph{Length of URL}
  \newline Attackers may use long URLs to mask the phishy appearance of phishing URLs. The length of URLs plays an important role in distinguishing phishing URLs from benign ones. We use the same length standards of ~\cite{DBLP:conf/icitst/MohammadTM12} as follows:
  \begin{align*}\begin{small}
  \textrm{A URL is }
  \begin{cases}
  \textrm{benign, if its length}\leq 53,\\
  \textrm{neutral, if } 54 \leq \textrm{its length} \leq 75,\\
  \textrm{phishy, if its length}\geq 76.
  \end{cases}
  \end{small}\end{align*}
  \item \emph{Presence of digits in domain}
  \newline Benign URLs do not have digits in the domain. The presence of digits in  the domain is a common characteristic of phishing URLs. We set this feature as true if any digits are encountered in the domain name part of the URL.
  
  \item \emph{Frequency of suspicious words}
  \newline We keep track of the frequencies of suspicious and most common words occurring in URLs. We choose several suspicious words like `confirm', `account', `signin', `update', `logon', `cmd', and `admin'. These words are selected after surveying the literature and real-world datasets including ours.
  
  \item \emph{Multiple sub-domains}
  \newline \cite{DBLP:conf/icitst/MohammadTM12} stated the criteria for classifying a URL as phishy based on the count of its sub-domains. If a URL's resource name part has more than three dots, then it is likely to be phishy. An example of such a URL is `http://www.outlook.3uwin.com'.
  
  \item \emph{Brand name modifications with `-'}
  \newline We downloaded the top-1000 most visited websites from Alexa and used them as popular brand names. Phishing URLs create similar names with prefixes or suffixes. For example, `microsoft-x.com' and `x-microsoft.com' are phishing URLs.
  
  \item \emph{Very long hostname}
  \newline Too long hostname typically indicates phishyness. If the length of a hostname is longer than 22, then it is phishy.
  
  \item \emph{Prefix or suffix separated by `-' to domain}
  \newline It is well known that phishing URLs tend to add prefixes or suffixes separated by `-' to their domain to lure users into believing that the website is legitimate. For instance, an attacker may use Amazon's domain separated by a prefix as `http://www.hello-amazon.com'.
  
  \item \emph{Frequency of punctuation symbols}
  \newline We count the occurrence of symbols like `.', `!', `\&', `,', `\#', `\$', and `\%'; ~\cite{Verma:2015:CPU:2699026.2699115} observed a high percentage of punctuation symbols in phishing URLs.

  \item \emph{The number of `:' in hostname}
  \newline  The number of `:' in the hostname part also implies phishyness. In particular, this is used for port number manipulation.

  \item \emph{Using Internet Protocol (IP) address}
  \newline Usage of IP addresses in place of domain names usually indicates fraudulent websites. For example, 
  
  `http://120.10.10.8/login.php' is most likely a phishing URL. Some attackers may use hexadecimal numbers in the domain part, \textit{e.g.,} `http://0x78.0xA.OxA.8'.
  
  \item \emph{Vowel/Consonant ratio in hostname}
  \newline This feature is to calculate the ratio of total vowels to total consonants in the hostname part. Phishing URLs do not follow the standard ratio.
  
  \item \emph{Very short hostname}
  \newline If hostname is very short (\textit{e.g.,} smaller than five), then it is an indicator of phishyness.
  
  \item \emph{Existence of `@' symbol}
  \newline Attackers can use `@' symbol to trick users by exploiting the property of browsers to ignore everything before `@' in the address bar. Attackers can use URLs such as 
  
  `http://www.google.com@atc.com' which causes the browser to ignore `www.google.com' and proceed to `atc.com'.
\end{enumerate}
\end{appendices}

\end{document}